\documentclass[a4paper,fleqn,usenatbib]{mnras}

\usepackage[T1]{fontenc}
\usepackage{ae,aecompl}

\usepackage{graphicx}	
\usepackage{amsmath}	
\usepackage{amssymb}	
\usepackage{subfigure}
\usepackage{multirow}
\usepackage{longtable}

\newcommand{\dinamo}{{\sc dinamo}}
\newcommand{\pcc}{\,{\rm cm}^{-3}}
\newcommand{\gcc}{\,{\rm g \, cm}^{-3}}

\newcommand{\um}{\, {\rm \mu m}}
\newcommand{\nm}{\, {\rm nm}}

\newcommand{\kel}{\, {\rm K}}

\newcommand{\msun}{\, {\rm M}_\odot}
\newcommand{\nh}{n_{\rm H}}
\newcommand{\nel}{n_{\rm e}}

\newcommand{\kpc}{\, {\rm kpc}}
\newcommand{\geleven}{G$11.2$-$0.3$}

\newcommand{\gtwentyseven}{G$27.4$+$0.0$}

\newcommand{\gtwentynine}{G$29.7$-$0.3$}

\newcommand{\kyr}{\, {\rm kyr}}

\newcommand{\chsq}{\chi_{\rm red.}^2}

\title[Dust properties in SNRs]{Properties of shocked dust grains in supernova remnants}

\author[Priestley et al.]{
  F. D. Priestley$^{1}$\thanks{Email: priestleyf@cardiff.ac.uk},
  H. Chawner$^{1}$,
  M. J. Barlow$^{2}$,
  I. De Looze$^{2,3}$,
  H. L. Gomez$^{1}$,
  \newauthor
  M. Matsuura$^{1}$
  \\
$^{1}$School of Physics and Astronomy, Cardiff University, Queen's Buildings, The Parade, Cardiff CF24 3AA, UK \\
$^{2}$Department of Physics and Astronomy, University College London, Gower Street, London WC1E 6BT, UK\\
$^{3}$Sterrenkundig Observatorium, Ghent University, Krijgslaan 281 - S9, 9000 Gent, Belgium\\
}

\date{Accepted XXX. Received YYY; in original form ZZZ}

\pubyear{2022}

\begin{document}
\label{firstpage}
\pagerange{\pageref{firstpage}--\pageref{lastpage}}
\maketitle

\begin{abstract}

  Shockwaves driven by supernovae both destroy dust and reprocess the surviving grains, greatly affecting the resulting dust properties of the interstellar medium (ISM). While these processes have been extensively studied theoretically, observational constraints are limited. We use physically-motivated models of dust emission to fit the infrared (IR) spectral energy distributions of seven Galactic supernova remnants, allowing us to determine the distribution of dust mass between diffuse and dense gas phases, and between large and small grain sizes. We find that the dense ($\sim 10^3 \pcc$), relatively cool ($\sim 10^3 \kel$) gas phase contains $>90\%$ of the dust mass, making the warm dust located in the X-ray emitting plasma ($\sim 1 \pcc$/$10^6 \kel$) a {negligible} fraction of the total, {despite dominating the mid-IR emission. {The ratio of small ($\lesssim 10 \nm$) to large ($\gtrsim 0.1 \um$) grains in the cold component is consistent with that in the ISM, and possibly even higher}, whereas the hot phase is almost entirely devoid of {small grains}. This suggests that grain shattering, which processes large grains into smaller ones, is {ineffective in the low-density gas, contrary to model predictions}.} Single-phase models of dust destruction in the ISM, {which do not account for the existence of the cold swept-up material containing most of the dust mass,} are likely to {greatly} overestimate the rate of dust destruction by supernovae.

\end{abstract}

\begin{keywords}

  dust, extinction -- ISM: supernova remnants -- ISM: evolution

\end{keywords}



\section{Introduction}

Core-collapse supernovae (CCSNe) both form \citep{dunne2009,matsuura2011,gomez2012,delooze2017,delooze2019,chawner2019,niculescu2021} and destroy \citep{jones1996,slavin2015} significant quantities of dust. The balance between these two processes determines whether CCSNe are net dust sources or sinks, which has important consequences for the evolution of the interstellar medium (ISM) \citep{morgan2003,delooze2020,galliano2021}. With extensive observational evidence for efficient dust formation by CCSNe, the key uncertainties in models of ISM evolution are now those related to dust destruction {in the shockwaves driven by these same objects.}

{Although theoretical predictions for the destruction of newly-formed dust in CCSN ejecta} span nearly the entire range from complete destruction to complete survival (\citealt{kirchschlager2019} and references therein), models of dust destruction in the ISM have settled on a typical gas mass `cleared' of dust of order $\sim 10^3 \msun$ {per SN} (\citealt{jones1996,slavin2015}, although \citealt{kirchschlager2022} report larger values). {This corresponds to} $\sim 10 \msun$ of dust destroyed per SN for a typical Galactic dust-to-gas (DTG) ratio of $\sim 0.01$, {much larger than the observed dust masses of $\lesssim 1 \msun$ found in supernova remnants (SNRs), suggesting that CCSNe are net destroyers of dust under present-day ISM conditions.}

These models {generally assume that} shocks propagate into a spherically symmetric, uniform density ISM, {with properties appropriate for the warm neutral medium ($\nh \sim 1 \pcc$, $T \sim 10^4 \kel$). The real ISM is multi-phase, with most of the mass concentrated in colder, denser gas ($\sim 30 \pcc$/$100 \kel$; \citealt{mckee1977}).} More realistic ISM structures can significantly reduce the quantity of dust destroyed \citep{hu2019,martinez2019}, as denser regions of the ISM experience less violent shock interactions. Models also typically assume the standard \citet{mathis1977} (MRN) grain size distribution for the ISM dust, but this may vary depending on the ISM phase \citep{hirashita2009} and can be altered by the SN itself \citep{hoang2019}, with potentially dramatic effects on the resulting mass of dust destroyed \citep{kirchschlager2019}.

Observationally, infrared (IR) studies of SNRs often find dust temperatures lower than those predicted by models of grains in the high-temperature ($\gtrsim 10^6 \kel$) gas {produced by SN shocks in low-density material \citep{seok2015,koo2016,chawner2020}. Using data extending into the far-IR, \citet{chawner2020b} showed that the observed spectral energy distribution (SED) of the Tornado SNR cannot be explained by dust grains in the hot, X-ray emitting shocked material swept up by the expanding SNR. The far-IR emission requires the presence of colder grains, located within the SNR but not exposed to the high-temperature gas. This cold dust component has a mass at least an order of magnitude larger than that of the warmer ($\sim 100 \kel$) grains located in the hot gas. The same phenomenon was found for the three SNRs investigated in \citet{priestley2021}, suggesting that it is not uncommon. Most observational studies of dust in SNRs \citep[e.g.][]{borkowski2006,williams2006,temim2012} have focused on the mid-IR emission, which is mainly produced by the warm grains, and thus may represent a very small fraction of the total dust mass within the SNR. The dust properties derived from the mid-IR data, and the resulting destruction efficiencies, are potentially unrepresentative of most of the swept-up material.}

{A significant issue in interpreting far-IR observations of SNRs is that the dust properties are poorly constrained; in \citet{chawner2020b} and \citet{priestley2021}, we were forced to assume standard ISM properties for both the cold dust, and for the radiation field presumably responsible for heating it. The impact of the SNR blastwave can be expected to significantly alter both of these properties. In this paper, we develop a comprehensive model for the IR emission of shocked dust grains in SNRs, allowing us to derive the properties of these grains from observational data, and thus provide empirical constraints on models of dust destruction.}

\begin{table*}
  \centering
  \caption{Distance, radius, {estimated age}, volume, the hot component density and temperature, the {initial} ISM density and shock velocity {reproducing these hot component properties}, and the scaling factor to reproduce observed X-ray luminosities, for each SNR in our sample. Note that for G11, G27 and G29, `volume' refers to {that of} the shell of shocked material, rather than the spherical volume. References: (1) \citet{priestley2021}; (2) \citet{chawner2020b}; (3) \citet{green2004}; (4) \citet{ranasinghe2018}; (5) \citet{verbiest2012}; (6) \citet{sawada2011}; (7) \citet{frail1996}; (8) \citet{kilpatrick2016}; (9) \citet{chawner2020}; (10) \citet{kothes2007}; (11) \citet{tian2014}; (12) \citet{keohane2007}; (13) \citet{park2010}; (14) \citet{leahy2020}; (15) \citet{koo2016}; (16) \citet{chawner2019}.}
  \begin{tabular}{ccccccccccc}
    \hline
    SNR & $D$/kpc & $r$/pc & Age/$\kyr$ & $V$/pc$^3$ & $\nh$/cm$^{-3}$ & $T$/$10^6 \kel$ & $n_{\rm ISM}$/cm$^{-3}$ & $v_{\rm sh}$/km s$^{-1}$ & $f$ & Ref. \\
    \hline
    G11 & $4.4$ & $2.8$ & $1.4-2.4$ & $33.7$ & $6.8$ & $8.2$ & $1.7$ & $770$ & $0.028$ & 1, 3, 15, 16 \\
    G27 & $5.8$ & $2.9$ & $0.8-2.1$ & $37.1$ &  $6.5$ & $9.1$ & $1.6$ & $810$ & $0.019$ & 1, 4, 15, 16 \\
    G29 & $5.8$ & $3.2$ & $<0.8$ & $59.4$ & $1.6$ & $26.0$ & $0.4$ & $1370$ & $0.010$ & 1, 5, 15, 16 \\
    Tornado & $11.8$ & $4.5$ & $2-8$ & $382$ & $0.5$ & $8.5$ & $0.13$ & $780$ & $0.091$  & 2, 6, 7 \\
    G43 & $11.3$ & $8.2$ & $1-4$ & $2310$ & $1.0$ & $18.6$ & $0.25$ & $1170$ & $0.121$ & 8, 9, 12, 15, 16 \\
    G340 & $15.0$ & $13.6$ & $2.6$ & $10530$ & $1.0$ & $11.6$ & $0.25$ & $920$ & $0.121$ & 9, 10, 13, 16 \\
    G349 & $11.5$ & $4.5$ & $1.8$ & $382$ & $3.4$ & $7.0$ & $0.85$ & $710$ & $0.091$ & 9, 11, 14, 15, 16 \\
    \hline
  \end{tabular}
  \label{tab:snrprop}
\end{table*}

\begin{figure*}
  \centering
  \includegraphics[height=\columnwidth]{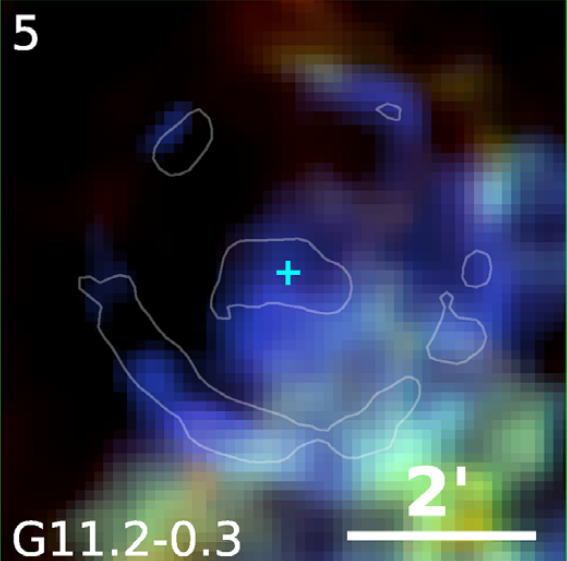}\hfill
  \includegraphics[height=\columnwidth]{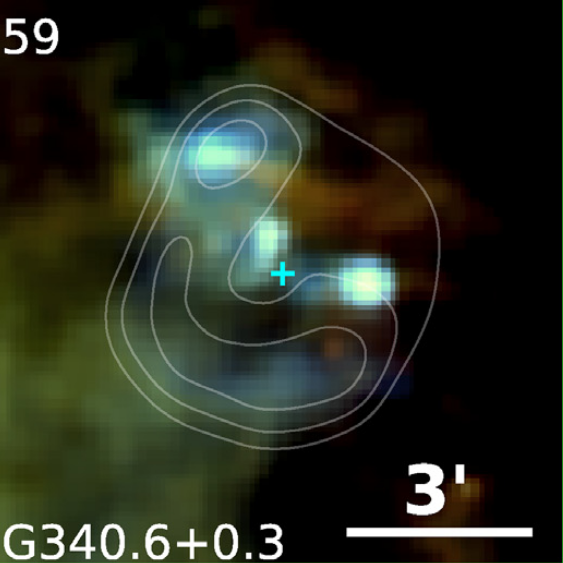}
  \caption{Far-IR {\it Herschel} three-colour images (red $250 \um$; green $160 \um$; blue $70 \um$) of G11 (left) and G340 (right), with X-ray flux overlaid as contours. Reproduced from \citet{chawner2019}.}
  \label{fig:img}
\end{figure*}

\begin{figure*}
  \centering
  \includegraphics[width=\columnwidth]{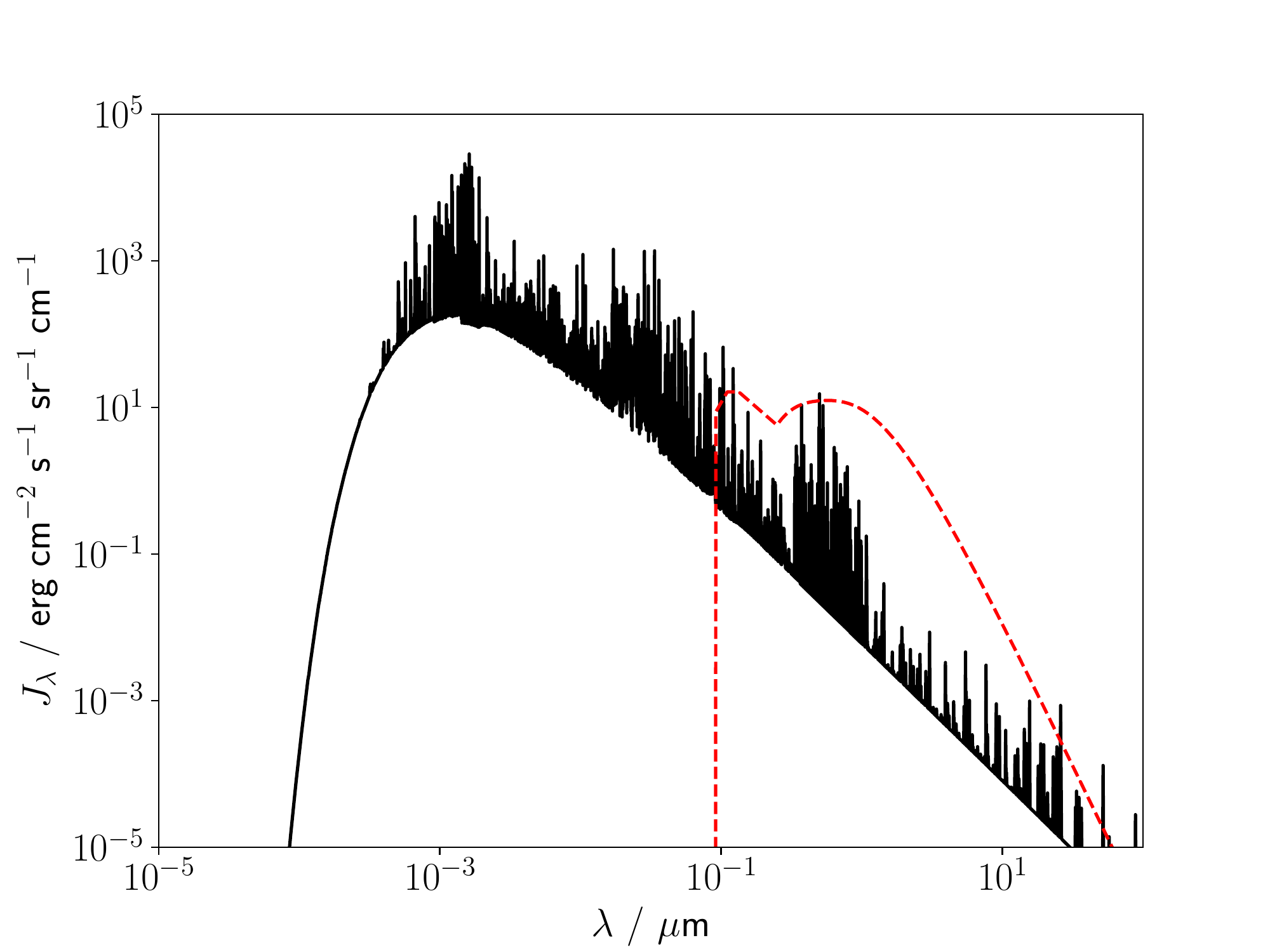}\quad
  \includegraphics[width=\columnwidth]{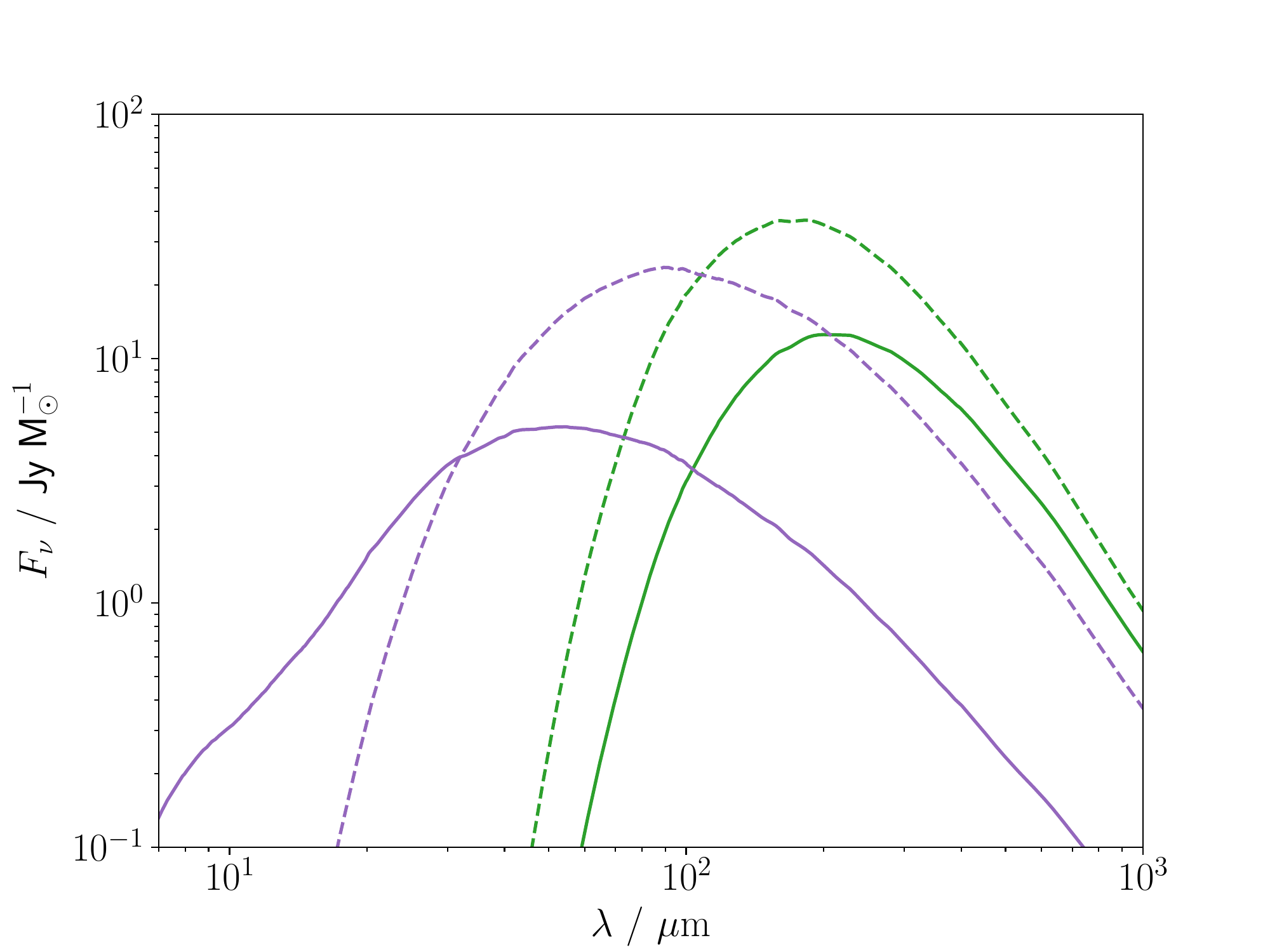}
  \caption{{\it Left:} shock-generated radiation field for G11 (black) compared to the \citet{mathis1983} ISM radiation field (red). {\it Right:} {flux per unit mass for $0.1 \um$ (green) and $5 \nm$ (purple) carbon grains, heated by the G11 (solid lines) and ISM (dashed lines) radiation fields.}}
  \label{fig:sed}
\end{figure*}

\begin{figure*}
  \centering
  \subfigure{\includegraphics[width=\columnwidth]{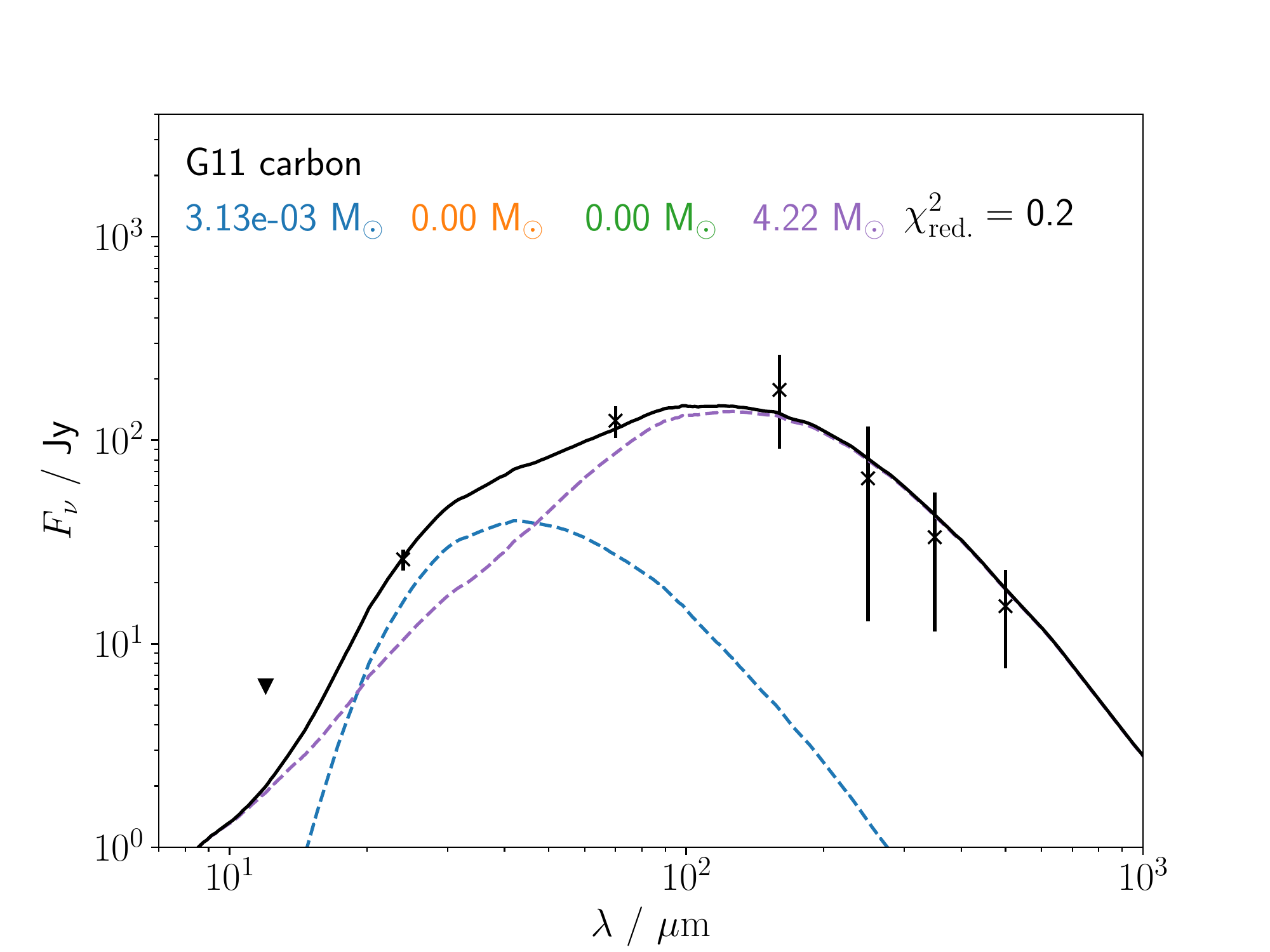}}\quad
  \subfigure{\includegraphics[width=\columnwidth]{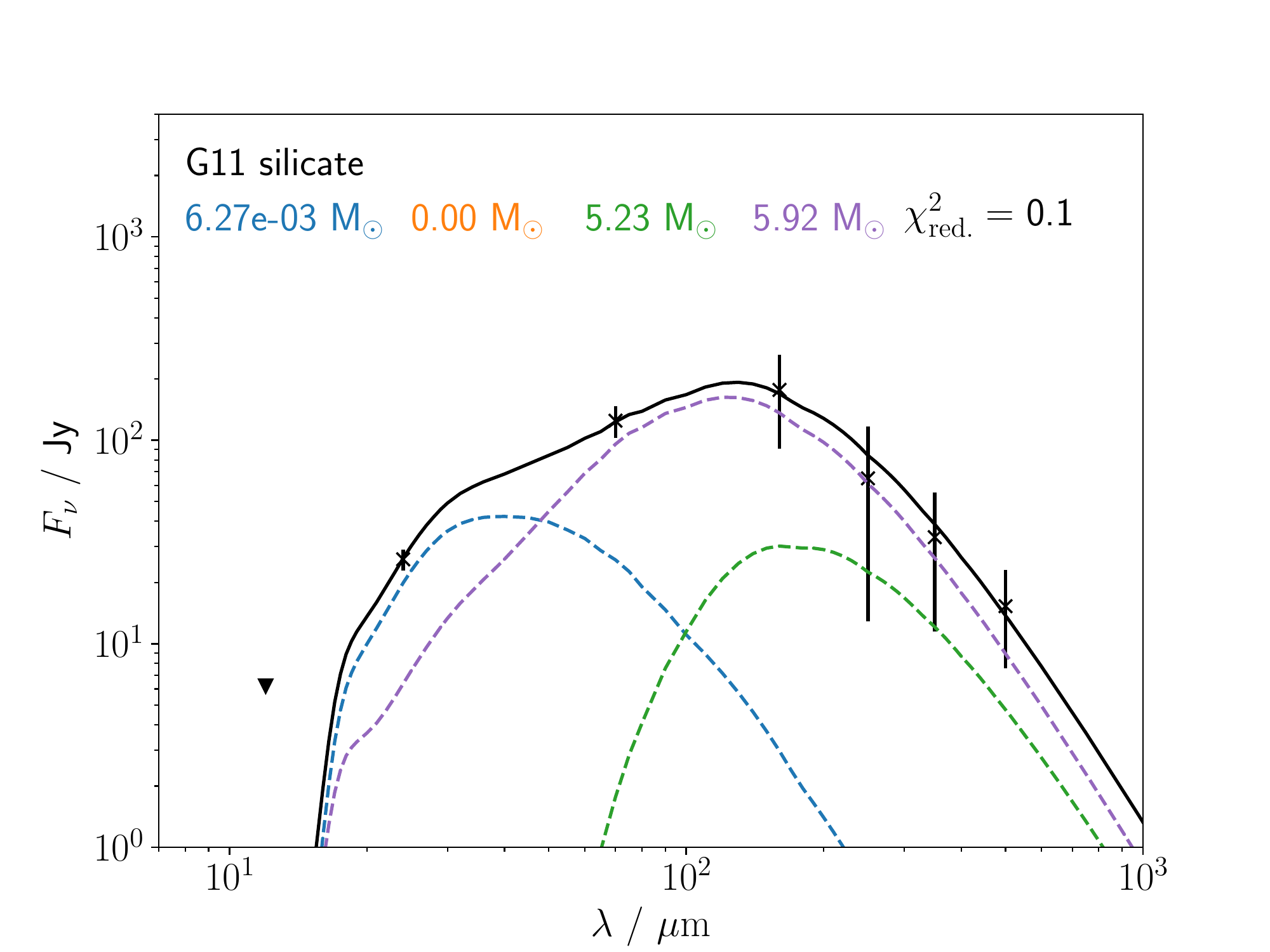}}
  \caption{Best-fit dust SEDs for G11 - data (black crosses), total model SED (black line), and individual component SEDs: hot/large (blue); hot/small (orange); cold/large (green); cold/small (purple). {Note that some component SEDs may not be visible, due to their contributing a negligible amount to the total SED.} {\it Left:} carbon grains. {\it Right:} silicates.}
  \label{fig:g11fit}
\end{figure*}

\begin{figure}
  \centering
  \subfigure{\includegraphics[width=\columnwidth]{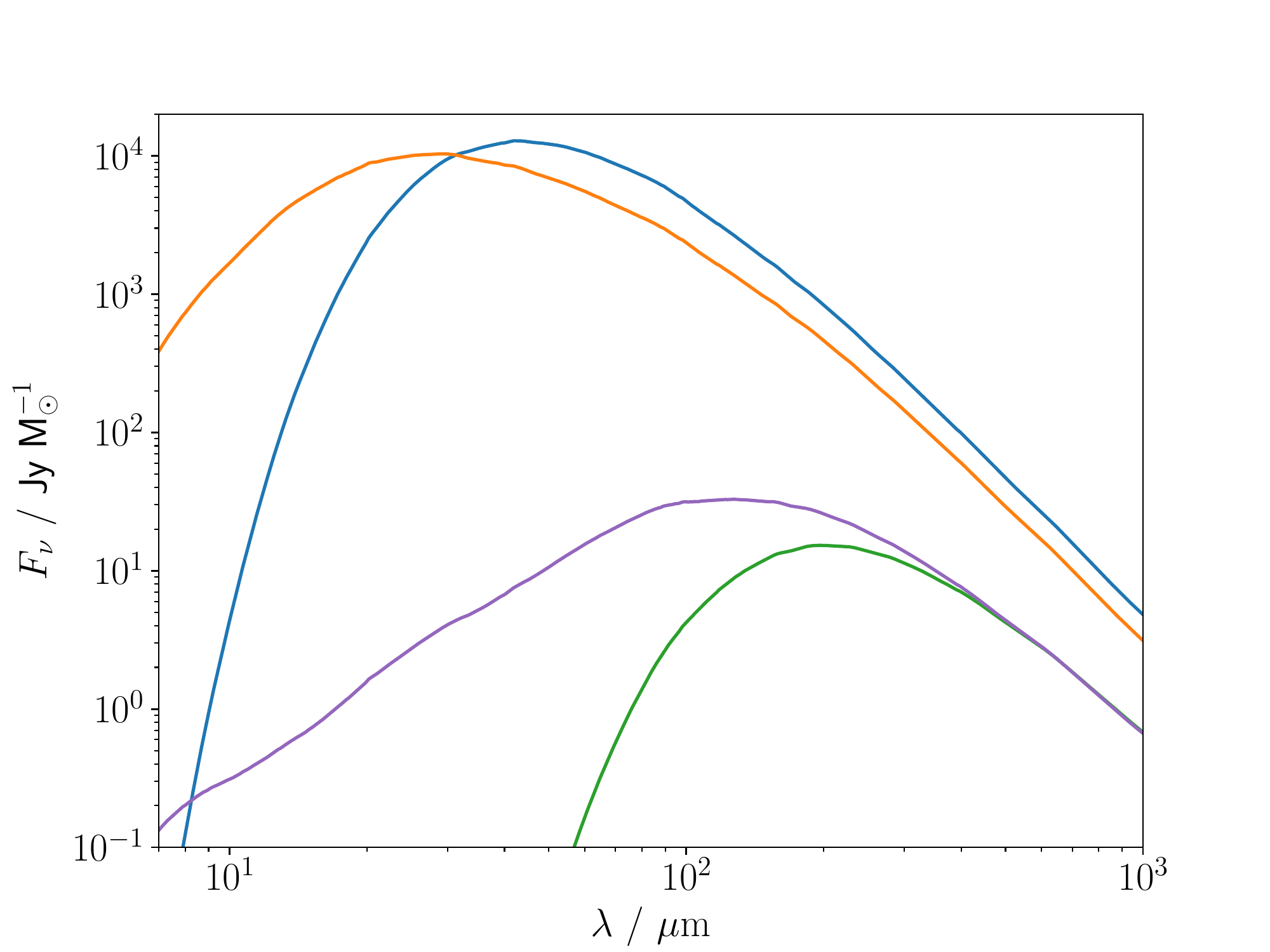}}\quad
  \caption{Flux per unit mass for carbon grains in G11: hot/large (blue); hot/small (orange); cold/large (green); cold/small (purple).}
  \label{fig:g11emis}
\end{figure}

\begin{figure*}
  \centering
  \subfigure{\includegraphics[width=\columnwidth]{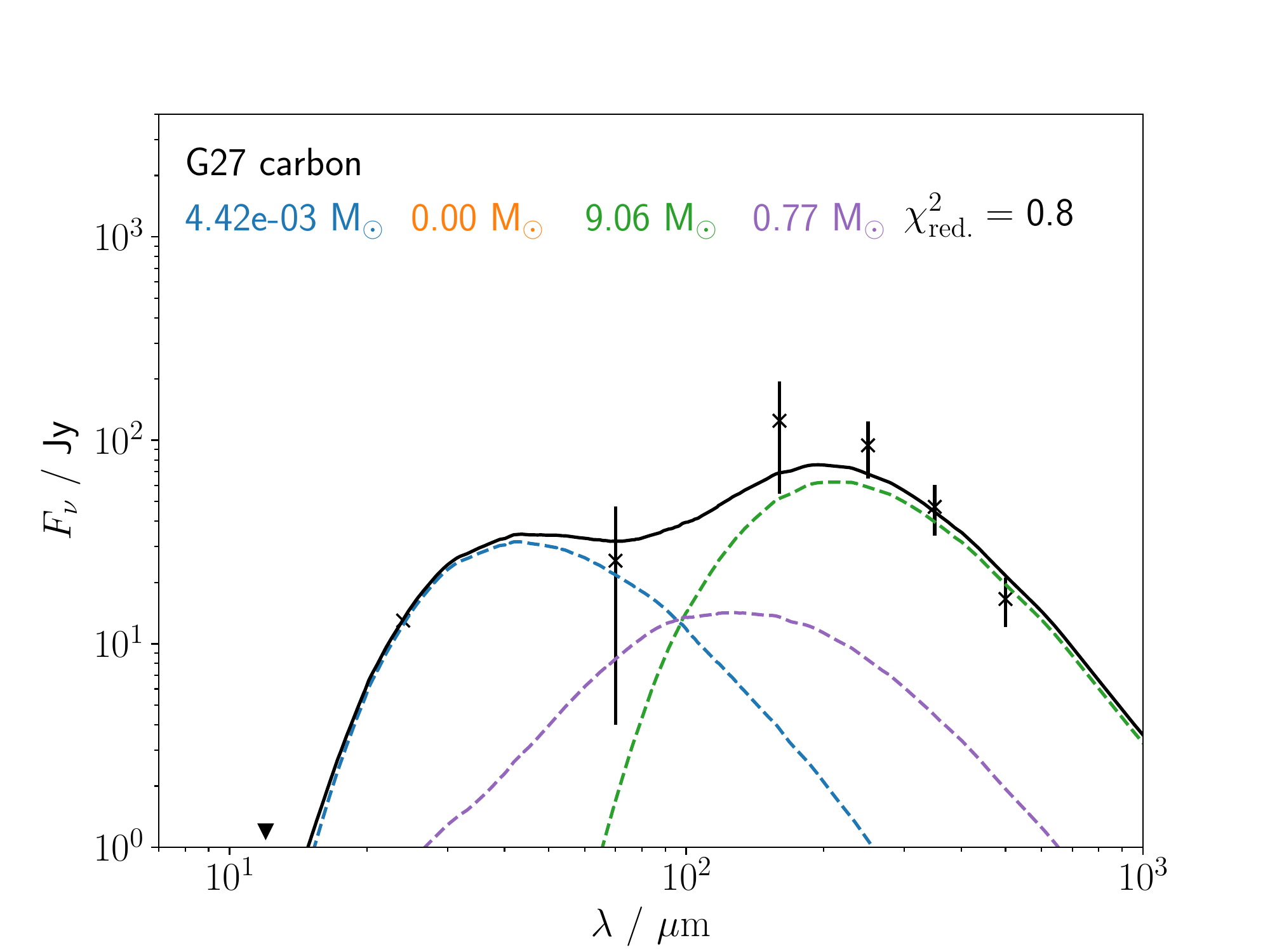}}\quad
  \subfigure{\includegraphics[width=\columnwidth]{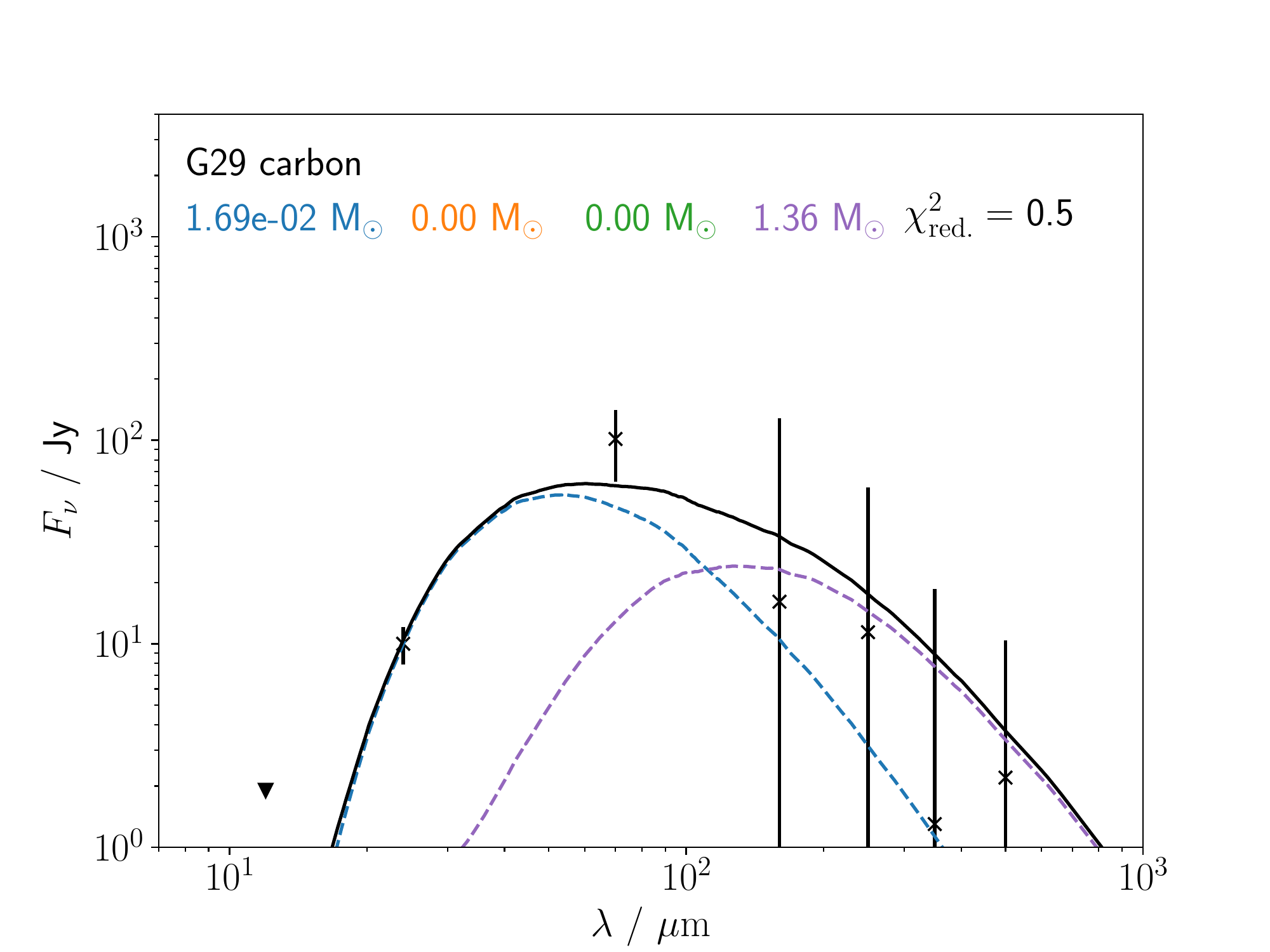}}\\
  \subfigure{\includegraphics[width=\columnwidth]{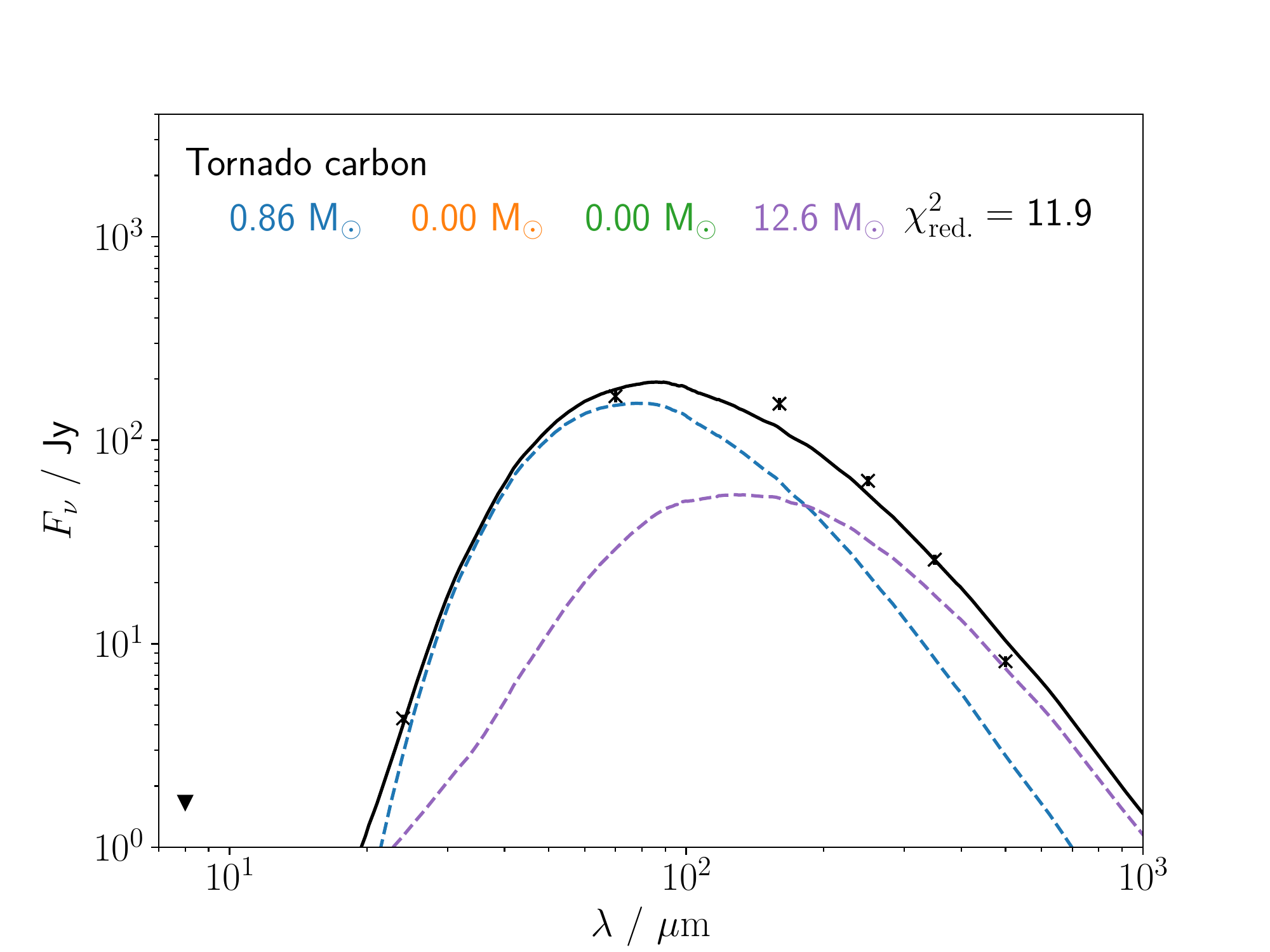}}\quad
  \subfigure{\includegraphics[width=\columnwidth]{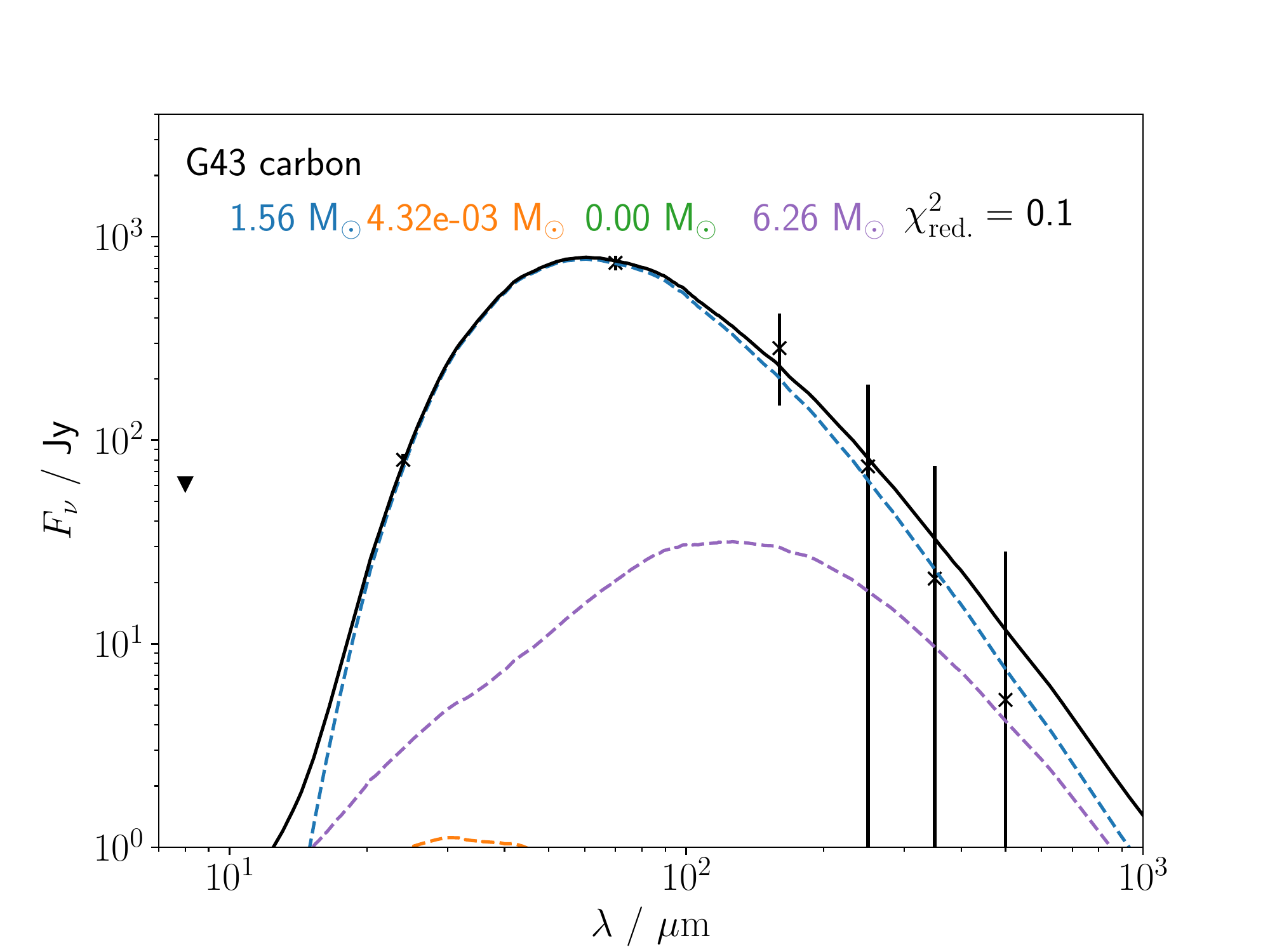}}\\
  \subfigure{\includegraphics[width=\columnwidth]{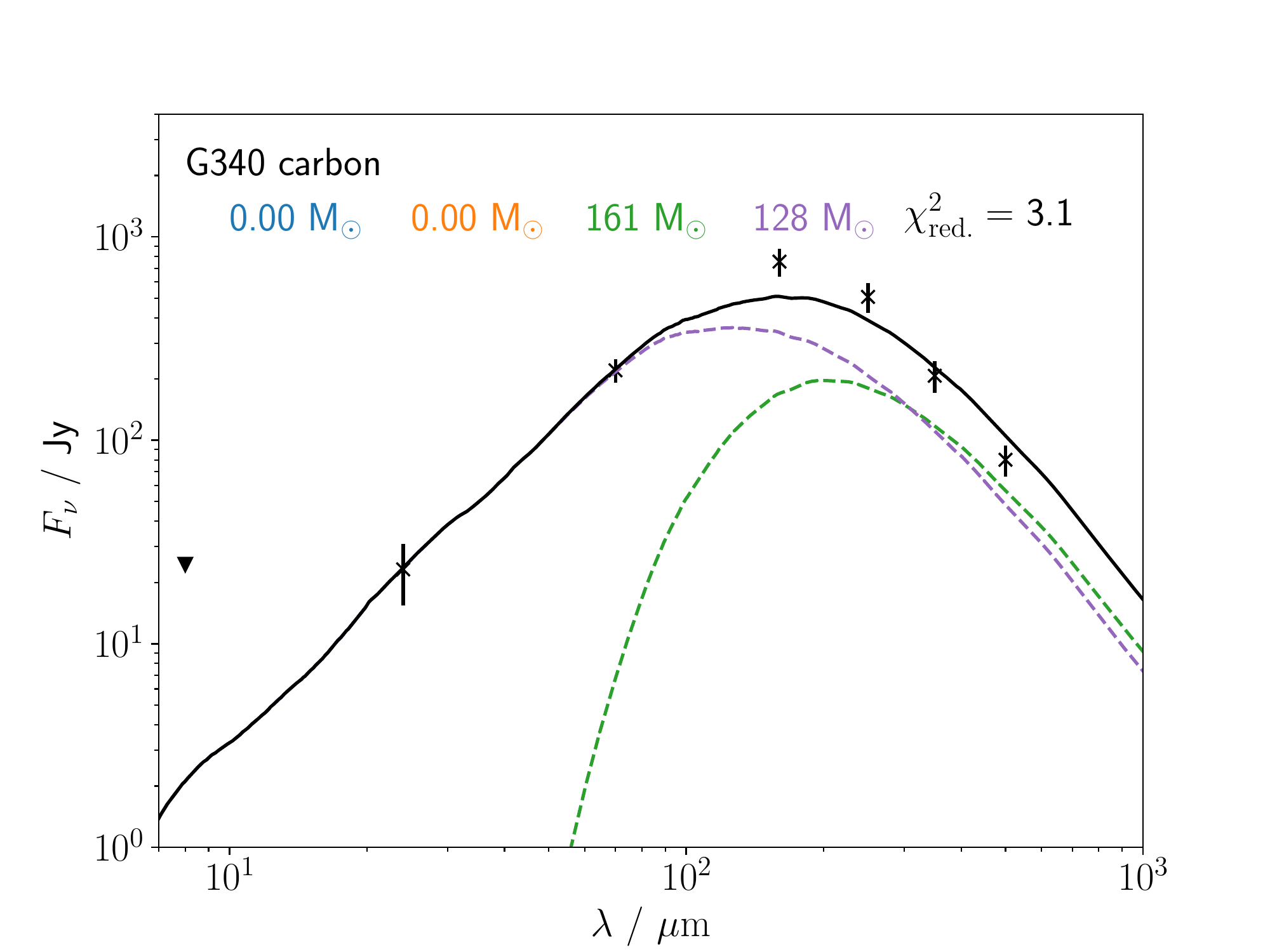}}\quad
  \subfigure{\includegraphics[width=\columnwidth]{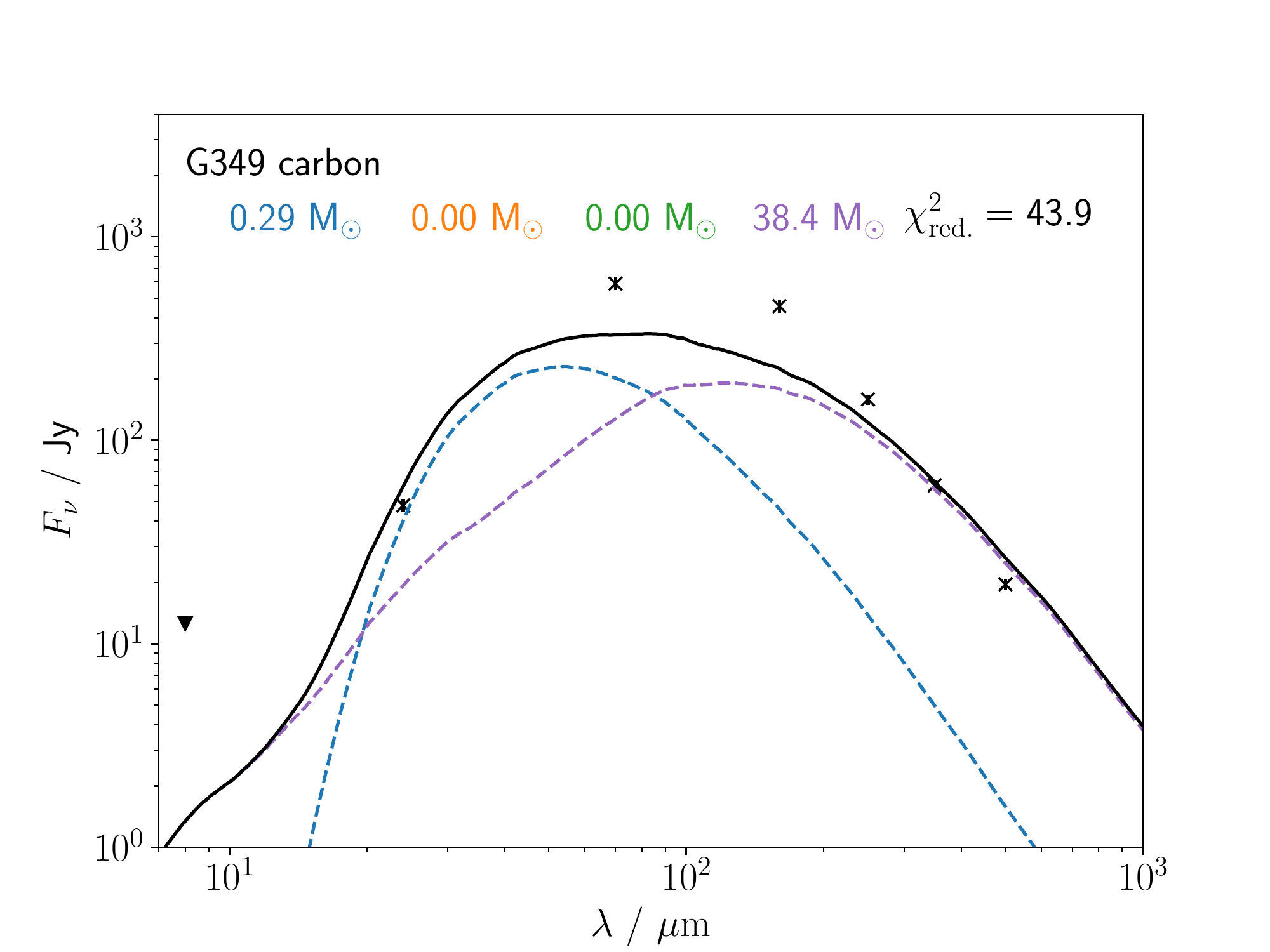}}\\
  \caption{Best-fit dust SEDs for carbon grains - data (black crosses), total model SED (black line), and individual component SEDs: hot/large (blue); hot/small (orange); cold/large (green); cold/small (purple). {\it Top left:} G27. {\it Top right:} G29. {\it Middle left:} Tornado. {\it Middle right:} G43. {\it Bottom left:} G340. {\it Bottom right:} G349.}
  \label{fig:allfits}
\end{figure*}

\begin{table*}
  \centering
  \caption{Median SNR dust masses returned by the MCMC for each component, with the 16th and 84th percentiles as uncertainties, {and the estimated initial swept-up dust mass $M_0$, assuming an initial density $n_{\rm ISM}$ and volume $V$ from Table \ref{tab:snrprop}, and a DTG mass ratio of $0.01$.}}
  \begin{tabular}{cccccc}
    \hline
    & \multicolumn{4}{c}{Dust mass / $\msun$} & \\
    SNR & Hot/large & Hot/small & Cold/large & Cold/small & $M_0/\msun$ \\
    \hline
    \multicolumn{6}{c}{Carbon} \\
    \hline
    G11 & $3.0^{+0.7}_{-0.8} \times 10^{-3}$ & $< 10^{-4}$ & $<0.2$ & $4.1^{+0.7}_{-0.8}$ & $0.020$ \\
    G27 & $4.6^{+0.1}_{-0.9} \times 10^{-3}$ & $< 10^{-4}$ & $9.6^{+1.3}_{-1.7}$ & $<0.5$ & $0.021$ \\
    G29 & $0.017^{+0.003}_{-0.003}$ & $< 2 \times 10^{-4}$ & $0.004^{+0.396}_{-0.003}$ & $<0.3$ & $0.008$ \\
    Tornado & $0.86^{+0.05}_{-0.04}$ & $< 10^{-4}$ & $<0.2$ & $12.5^{+0.6}_{-0.9}$ & $0.017$ \\
    G43 & $1.61^{+0.07}_{-0.09}$ & $< 5 \times 10^{-3}$ & $<2$ & $<1$ & $0.203$ \\
    G340 & $<0.7$ & $< 4 \times 10^{-3}$ & $167^{+98}_{-31}$ & $123^{+14}_{-99}$ & $0.928$ \\
    G349 & $0.30^{+0.03}_{-0.02}$ & $< 10^{-4}$ & $<0.5$ & $38.2^{+1.2}_{-1.8}$ & $0.114$ \\
    \hline
    \multicolumn{6}{c}{Silicate} \\
    \hline
    G11 & $5.3^{+1.3}_{-5.0} \times 10^{-3}$ & $< 10^{-3}$ & $<4$ & $6.6^{+1.0}_{-1.2}$ & $0.020$ \\
    G27 & $6.4^{+0.1}_{-2.5} \times 10^{-3}$ & $< 6 \times 10^{-4}$ & $40.5^{+5.0}_{-6.4}$ & $<0.6$ & $0.021$ \\
    G29 & $0.014^{+0.003}_{-0.004}$ & $< 1 \times 10^{-3}$ & $<1$ & $<4$ & $0.008$ \\
    Tornado & $0.307^{+0.020}_{-0.019}$ & $< 10^{-4}$ & $<0.5$ & $45.1^{+1.1}_{-1.3}$ & $0.017$ \\
    G43 & $0.47^{+0.10}_{-0.10}$ & $< 10^{-3}$ & $<2$ & $142^{+20}_{-23}$ & $0.203$ \\
    G340 & $<0.5$ & $< 0.03$ & $962^{+171}_{-117}$ & $130^{+25}_{-71}$ & $0.928$ \\
    G349 & $0.22^{+0.02}_{-0.02}$ & $< 2 \times 10^{-4}$ & $<0.5$ & $101.6^{+2.4}_{-2.8}$ & $0.114$ \\
    \hline
  \end{tabular}
  \label{tab:mass}
\end{table*}

\begin{figure}
  \centering
  \includegraphics[width=\columnwidth]{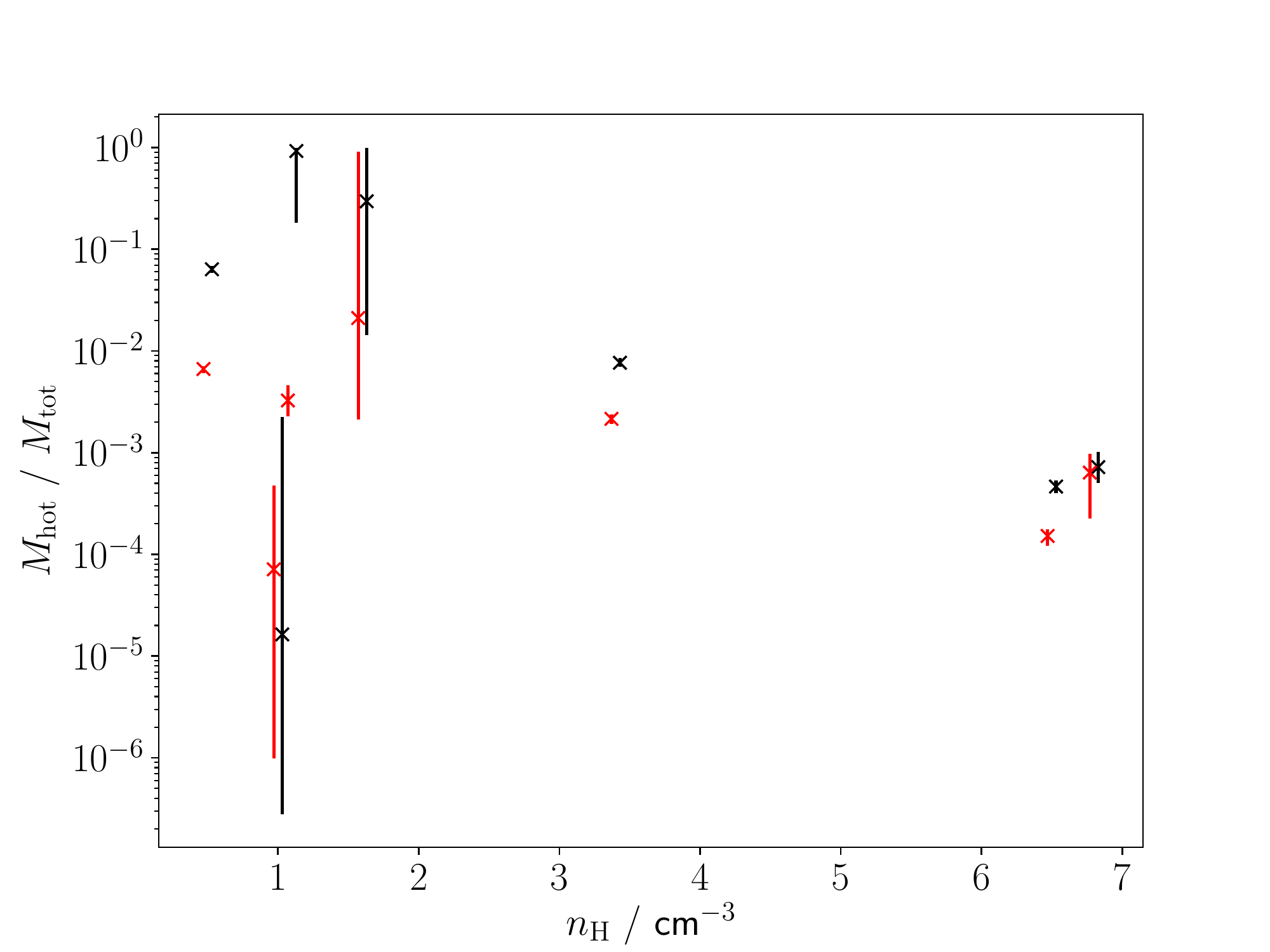}
  \caption{Ratio of the dust mass in the hot component {to the total dust mass} versus {hot component gas} density. Crosses represent median values of the MCMC, {error bars the 16th and 84th percentiles}. Values for carbon and silicate grains are shown in black and red respectively. G43 has been plotted at $\nh = 1.1 \pcc$ to avoid confusion with G340, {and the carbon and silicate data slightly offset from each other}.}
  \label{fig:massratio}
\end{figure}

\begin{figure*}
  \centering
  \includegraphics[width=\columnwidth]{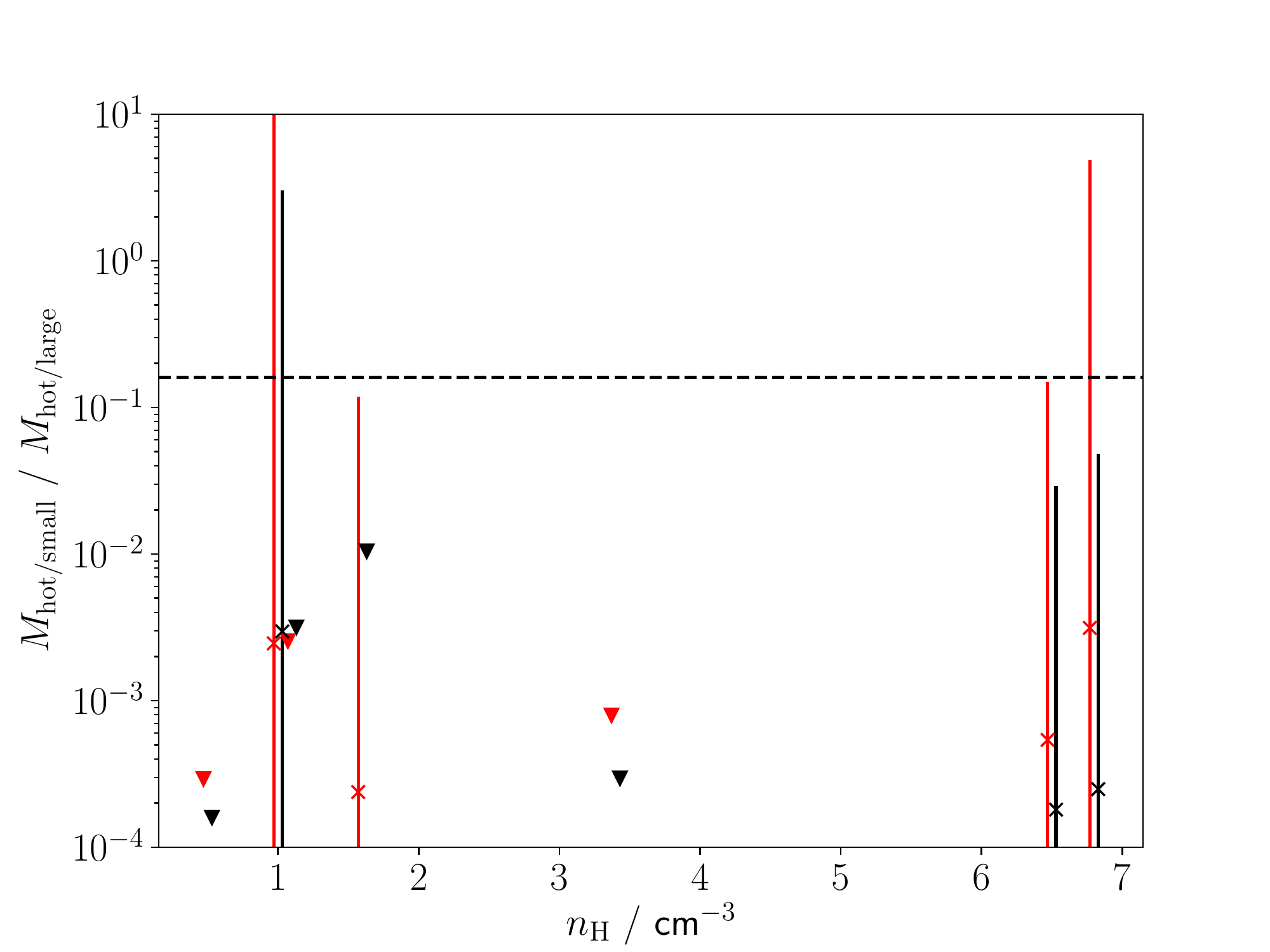}\quad
  \includegraphics[width=\columnwidth]{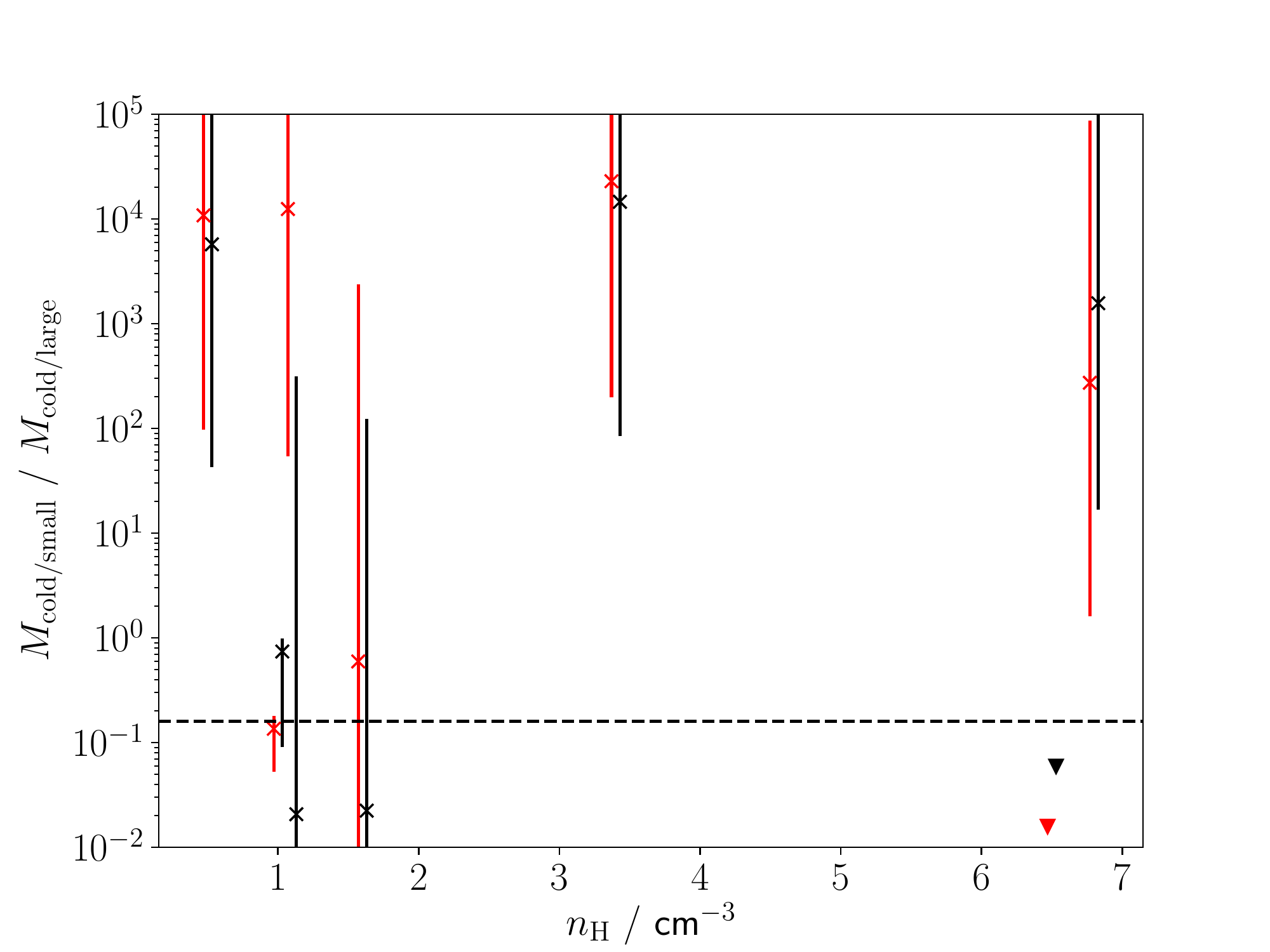}
  \caption{Ratio of the dust mass in small and large grains versus {hot component gas} density, in the hot (left) and cold (right) components. Crosses represent median values of the MCMC, {error bars the 16th and 84th percentiles, and} triangles the 84th percentile as an upper limit. Values for carbon and silicate grains are shown in black and red respectively. G43 has been plotted at $\nh = 1.1 \pcc$ to avoid confusion with G340, {and the carbon and silicate data slightly offset from each other}. An MRN small/large ratio of $0.16$ is marked with a dashed line.}
  \label{fig:sizeratio}
\end{figure*}

\section{Method}

\subsection{Observational sample}

We consider a sample of seven core-collapse SNRs: \geleven{} (hereafter G11), \gtwentyseven{} (G27), \gtwentynine{} (G29), G$43.3$-$0.2$ (G43), G$340.6$+$0.3$ (G340), G$349.7$+$0.2$ (G349), and G$357.7$-$0.1$ (the Tornado). {We have previously investigated the Tornado in \citet{chawner2020b}, and G11, G27 and G29 in \citet{priestley2021}. In all four SNRs, the IR SEDs were inconsistent with collisionally-heated dust grains, for gas properties derived from X-ray data. We found that an additional, colder dust component is required to reproduce the far-IR SEDs, but were unable to constrain its properties beyond estimating the dust masses involved.}

The other three SNRs (G43, G340 and G349) were selected from the \citet{chawner2019,chawner2020} catalogue of Galactic plane SNRs due to having both clear SNR-associated IR emission {in at least one band (most often {\it Spitzer} MIPS $24 \um$)}, and X-ray emission indicating ongoing shock interation with the surrounding ISM. These were excluded from the analysis in \citet{priestley2021} due to their irregular morphologies; {Figure \ref{fig:img} shows far-IR three-colour images of G11 and G340, with X-ray contours overlaid. While G11 has a coincident shell-like structure in both the far-IR and X-ray data, indicating an interaction between the SNR and the surrounding ISM, G340 is much more confused. Nonetheless, the combination of enhanced X-ray emission and dust temperature (represented by the $70 \um$ flux) within the SNR strongly suggests interaction with ambient material.} Table \ref{tab:snrprop} lists relevant physical properties for the SNRs.

{G11 and G29 both show evidence of newly-formed ejecta dust interior to the region of interaction with the ISM, found in the pulsar wind nebulae located at the centres of these SNRs\footnote{{In both cases, the radiation field generated by the pulsar wind nebula is insufficient to power the central dust emission \citep{priestley2020}. As the swept-up dust of interest is located at even greater distances,  we assume the impact of the central object is negligble.}} \citep{chawner2019}. For these objects, we take IR fluxes from \citet{priestley2021}, extracted from annuli excluding the central regions, and thus presumably dominated by the swept-up ISM. G27 has little central IR flux, but does have a shell-like X-ray structure, so we again use the \citet{priestley2021} annulus fluxes. Fluxes for the Tornado are taken from \citet{chawner2020b}, and those for G43, G340 and G349 from \citet{chawner2020}, using circular apertures. These last four SNRs have previously-derived dust masses ($\gg 1 \msun$) far in excess of what could be produced by a single CCSN, so we assume the IR fluxes are primarily due to ISM dust, and that any contribution from ejecta dust is negligble. The IR data for each SNR are listed in Appendix \ref{sec:snrsed}.}

\subsection{Dust SED model}

For each SNR, we calculate dust emission models using \dinamo{} \citep{priestley2019}, which determines the temperature distribution for grains heated by the local radiation field and electron/ion collisions. We assume there are two gas components in each SNR; a `hot' component responsible for the X-ray emission, with typical densities of $\sim 1 \pcc$ and temperatures $\gtrsim 10^6 \kel$, and an additional `cold' component with a much higher pre-shock density ($\gtrsim 100 \pcc$) and thus a much lower post-shock temperature ($\ll 10^6 \kel$). {The two components are assumed to be spatially well-mixed (e.g. cold clumps embedded in a hot diffuse medium) , so that the local radiation field is the same for both.}

The gas properties of the hot component are taken from analysis of the X-ray data in the literature, listed in Table \ref{tab:snrprop}, and we assume $\nel = \nh$ for simplicity. Properties for G11, G27 and G29 were derived from modelling X-ray data in \citet{priestley2021}, and those for the Tornado obtained similarly by \citet{sawada2011}. The temperature for G43 is taken from \citet{keohane2007}, and we use the lower end of their quoted range of densities ($1-3.5 \pcc$). \citet{park2010} report a temperature for G340 of $1-1.5 \, {\rm keV}$; we again take the lower limit, and assume a density of $1 \pcc$, typical for the rest of our sample. \citet{leahy2020} provide the temperature and the emission measure for G349, again from X-ray modelling, and we obtain the density from the SNR volume, the emission measure, and the assumption $\nh = \nel = {\rm constant}$ throughout the SNR.

For the cold component, we assume $\nh = 1000 \pcc$, $\nel = 0.1 \pcc$ and $T = 5000 \kel$ for all SNRs, as determined for G11 from fits to the \citet{andersen2011} H$_2$ line observations in \citet{priestley2021}. These properties may not be appropriate for the rest of our sample, but they are fairly typical values for SNRs interacting with dense ambient material, as derived from molecular line observations \citep[e.g.][]{reach2005,zhu2014}. {We discuss the importance of this assumption in Appendix \ref{sec:coldprop}.}

We assume the local radiation field {in both the hot and cold component} is primarily due to emission from the shock-ISM interaction. We use {\sc mappings} \citep{sutherland2017} to calculate plane-parallel radiative shock models, with the {initial gas density, $n_{\rm ISM}$, and shock velocity, $v_{\rm sh}$, chosen to reproduce the post-shock density and temperature reported by X-ray studies of each object} (the initial temperature is fixed to $10^4 \kel$). We then scale the resulting SED by a factor $f$, such that the X-ray luminosity of a spherical shell of emitting material, with the same radius as the SNR, matches the observed values given by \citet{koo2016}. The Tornado and G340 are not included in \citet{koo2016}, so we adopt the values of G349 and G43 respectively, being the best-matched SNRs in terms of {X-ray derived gas properties.}

For a spherical shell of emitting material, the flux at any point within the shell is the same as the flux at the centre by symmetry. We approximate the local radiation field {heating the dust} using this central radiation field. {While swept-up grains located at the shock front are much closer to {\it some} of the shocked ISM generating this radiation field, they are also much further away from most of it. The $r^{-2}$ scaling of the received flux with distance to the source should, roughly, cancel out the $\sim r^2$ growth in the amount of material emitting at a given distance. The typical radiation field experienced by a dust grain should therefore be comparable to that at the centre. While several of the SNRs in our sample clearly deviate from spherical symmetry,} this represents a substantial upgrade on our previous work \citep{chawner2020b,priestley2021}, where we assumed a \citet{mathis1983} ISM field scaled by an arbitrary constant.

Typical shock-generated radiation fields are very different from those in the wider ISM, {affecting the resulting dust SEDs. Figure \ref{fig:sed} shows the radiation field produced by this method for G11 compared to the \citet{mathis1983} ISM field, and corresponding SEDs for carbon grains of different sizes heated by the two fields. The higher ultraviolet (UV) flux from the ISM field results in $0.1 \um$ carbon grains being heated to higher temperatures than for the G11 field, whereas the much greater flux of X-ray photons from the SNR causes non-equilibrium effects in $5 \nm$ grains to become more important, reflected in the increased mid-IR grain emission. These effects are not necessarily universal, even for different grain types in the same radiation field (large silicate grains have similar temperatures for both the G11 and ISM fields) - it is essential to consider the local radiation field on a case-by-case basis.} Shock parameters and scaling factors are listed in Table \ref{tab:snrprop}.

We calculate dust SEDs for $0.1 \um$ and $5 \nm$ grains, representing the largest and smallest sizes typically present in the ISM \citep{mathis1977}, for the hot and cold components of each SNR. {We demonstrate in Appendix \ref{sec:sizedist} that our results are not sensitive to the specific choice of `large' and `small' grain sizes.} We then fit the observed IR SEDs\footnote{Flux measurements shortwards of $24 \um$ are treated as upper limits, due to the potential for significant non-dust contamination at these wavelengths.}, after convolving with the appropriate filter response curves, with the mass of each dust component (small/large, hot/cold) as the four free parameters. {In the following, we depict the hot/large dust component in blue, hot/small in orange, cold/large in green, and cold/small in purple.} Fitting is done using {\it emcee}, a Monte Carlo Markov chain (MCMC) code \citep{foreman2013}, with $500$ walkers, $5000$ steps per walker, and $500$ burn-in steps, which is sufficient for convergence. For dust properties, we use either carbon grains, with optical constants taken from \citet{zubko1996} and a bulk density of $1.6 \gcc$, or silicates, with MgSiO$_3$ optical constants from \citet{dorschner1995} (extended to far-UV/X-ray wavelengths with values from \citealt{laor1993}) and a bulk density of $2.5 \gcc$. While ISM dust includes both species, the data are insufficient to fit both simultaneously, and {our main conclusions hold regardless of the assumed grain composition}.

\section{Results}

Figure \ref{fig:g11fit} shows the results of our SED fitting for G11. The cold dust mass is at least three orders of magnitude larger than that of the hot component, consistent with the results from \citet{priestley2021}. We also find strict limits on the mass of small grains that can be present in the hot component, due to their high emissivity around $\sim 10 \um$ (Figure \ref{fig:g11emis}) combined with strong upper limits on the observed flux in this wavelength region, again consistent with our previous results. For either grain composition, we find that a significant mass of small grains are required in the cold component, at least comparable to that in large grains, and possibly much larger. This is in contrast to the MRN size distribution we assumed for this component in \citet{priestley2021}, where most of the mass is in the largest grain sizes, and suggests significant dust processing in the shocked material.

Figure \ref{fig:allfits} shows the best-fit carbon grain SEDs for the remaining six SNRs in our sample (results for silicate grains are shown in Appendix \ref{sec:silfit}). While the total dust mass in each SNR ranges from $\sim 1-100 \msun$, in all cases this mass is primarily in the cold component, with hot grains {typically} making up a negligible fraction of the total. The hot component consists only of large grains in all SNRs {for which it has a non-negligible mass}, and for all but one (G27), small grains make up a substantial fraction of the dust mass in the cold component. The parameter distributions returned by the MCMC confirm these findings. Median masses for each dust component\footnote{{These can differ significantly from the best-fit masses in Figures \ref{fig:g11fit} and \ref{fig:allfits}, generally indicating that the mass of that component is poorly constrained due to large observational uncertainties (e.g. G29).}} and the 16th and 84th percentiles, listed in Table \ref{tab:mass}, show that for all seven SNRs, the hot/small dust mass is consistent with zero, and typically constrained to be much lower than the other three dust components, regardless of the assumed grain composition.

{The total SNR dust masses for G11, G27, G29 and the Tornado are all within a factor of a few of our previous estimates for these objects \citep{chawner2020b,priestley2021}. For G340 and G349, our masses for carbon grains are in good agreement with those obtained from blackbody fits by \citet{chawner2020}, while with silicate optical properties the dust masses are somewhat larger, due to the lower mass opacity. Our carbon mass for G43 is significantly lower than the \citet{chawner2020} estimate because in this case, the far-IR fluxes can be reproduced by grains in the hot component, with high grain temperature and thus high emissivity. With silicate grains, for which this is not the case, our estimated mass is a factor of a few larger, as with G340 and G349. In general, our model results in dust masses basically consistent with those from previous work. The large ($> 100 \msun$) dust masses found for some of the SNRs are required by the observed far-IR fluxes, for the assumed distances in Table \ref{tab:snrprop} and typical dust mass opacities at these wavelengths.}

{We estimate the mass of swept-up ISM in the SNRs using the ambient densities derived from the X-ray data, $n_{\rm ISM}$, and volumes corresponding to the regions the IR SEDs were extracted from, $V$, as given in Table \ref{tab:snrprop}. Assuming an ISM DTG ratio of $0.01$ {and a hydrogen mass fraction of $0.7$}, the total swept-up dust masses for each SNR are given in Table \ref{tab:mass}, if the average density of the ambient ISM is that of the (pre-shock) hot component. These values are {all $<0.1 \msun$. The total observed dust mass exceeds our estimate for all seven SNRs, and by factors of $\gtrsim 100$ for all SNRs except G29. In order to contain this much dust, the average ambient ISM density around the SNRs would have to be larger than the $\sim 1 \pcc$ values in Table \ref{tab:snrprop} by a similar factor, i.e. $\left<n_{\rm ISM}\right> \gtrsim 100 \pcc$, comparable to the typical average densities of molecular clouds on these scales \citep{larson1981}.}

Figure \ref{fig:massratio} shows the ratio of the {hot component dust mass to the total dust mass} for each SNR, plotted against the density of the hot component. With a few exceptions, this value is well-constrained to be $< 0.1$, and in most cases $<0.01$ - the dust located in the high-temperature shocked material makes up a percent-level fraction of the total swept-up dust mass. {Due to the higher grain temperatures, this dust component typically contributes most of the mid-IR flux, but it is unrepresentative of the bulk of the swept-up dust mass.} {There is a suggestion of a negative correlation between the hot component dust mass fraction and the gas density. This could indicate either an intrinsically lower mass fraction of this component in denser regions, or more efficient sputtering in denser gas destroying a larger proportion of the initial dust mass.}

Figure \ref{fig:sizeratio} shows the mass ratio of small to large grains in both the hot and the cold components, again plotted against density. The mass ratio of grains with radii $\le 10 \nm$ to those $\ge 0.1 \um$ for an MRN size distribution ($\sim 16 \%$) is indicated for comparison. In general, it appears that the hot component is substantially depleted in small grains compared to the undisturbed ISM, whereas the cold component either has a more typical size distribution, or is enhanced in small grains, {to the extent that the mass of large grains is negligible in some SNRs} (note that because grain mass is proportional to $a^3$, the enhancement in grain number is even larger than indicated by Figure \ref{fig:sizeratio}).

\section{Discussion}

\begin{figure*}
  \centering
  \subfigure{\includegraphics[width=\columnwidth]{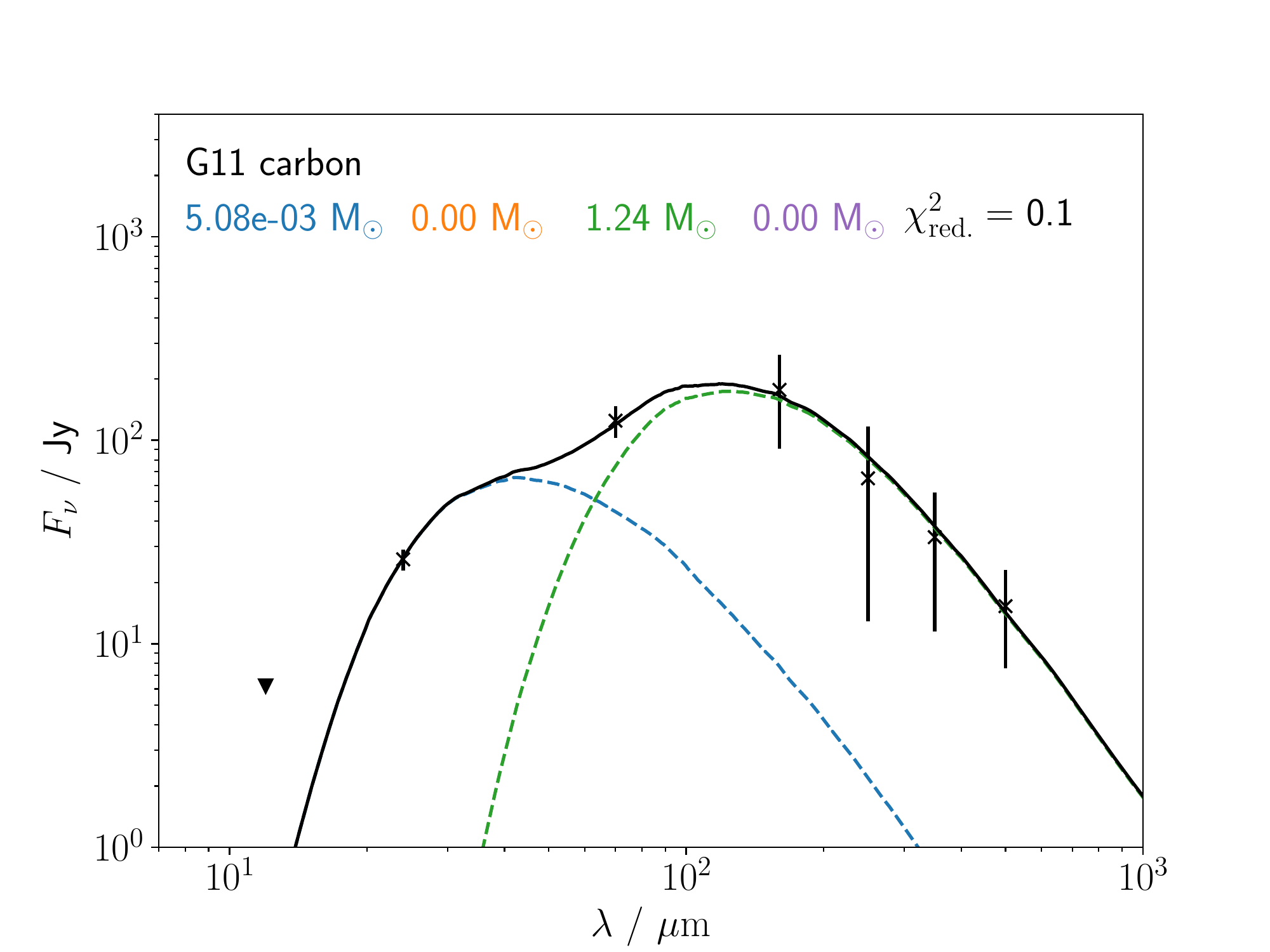}}\quad
  \subfigure{\includegraphics[width=\columnwidth]{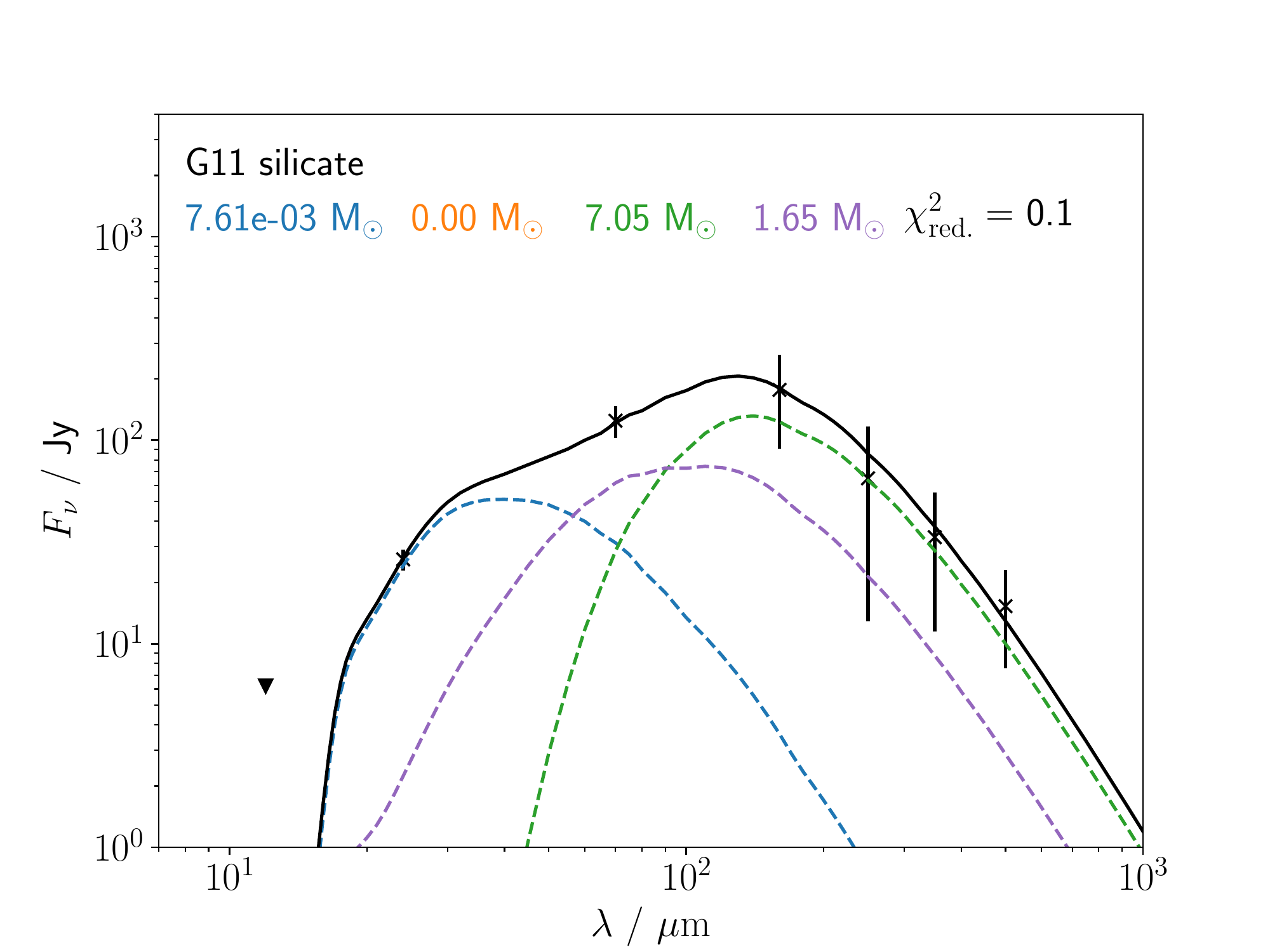}}
  \caption{Best-fit dust SEDs for G11, {with radiative heating by both the shock-generated radiation field, and a \citet{mathis1983} ISM field enhanced by a factor of five} - data (black crosses), total model SED (black line), and individual component SEDs: hot/large (blue); hot/small (orange); cold/large (green); cold/small (purple). {Note that some component SEDs may not be visible, due to their contributing a negligible amount to the total SED.} {\it Left:} carbon grains. {\it Right:} silicates.}
  \label{fig:g11ref}
\end{figure*}

\subsection{Caveats}

{The flux measurements we model (Appendix \ref{sec:snrsed}) are not the raw observational fluxes, but the estimated contribution from the SNRs at each wavelength, after the subtraction of background flux from unrelated material. There is no clear consensus on the best way to estimate these background contributions, and even if they are explicitly included as an additional model component, some assumptions have to be made about its properties (such as SED shape; see \citealt{delooze2017,delooze2019}). Given this freedom of choice, it is impossible to rigorously account for the possible impact on our results. We have thus refrained from drawing any conclusions based on single objects in our sample. The consistent qualitative results seen in Figures \ref{fig:massratio} and \ref{fig:sizeratio} across the sample of SNRs suggest that these are real physical phenomena, even if the numerical values for any individual SNR should be viewed with some caution.}

{Our sample was selected from the \citet{chawner2019,chawner2020} SNR catalogue based on the presence of co-spatial X-ray and warm dust ($24/70 \um$) emission, taken as a sign of interaction. We are therefore biased towards SNRs which are detectable in both these shock tracers, which likely corresponds to high ambient densities. In fact, all seven of our SNRs show signs of interaction with molecular material (G11, G27, G29, G43 - \citealt{kilpatrick2016}; G340 - \citealt{green1997}; G349 - \citealt{lazendic2010}; Tornado - \citealt{hewitt2008}). These interactions are not rare; $\sim 20 \%$ ($88/383$) of the SNRs in the \citet{ferrand2012} catalogue are listed as interacting with molecular clouds, and this seems likely to be an underestimate given the requirement for targeted molecular line observations. While our sample may well be biased, we do not appear to be selecting for a particularly uncommon class of object.}

The best-fit models for several SNRs (Figures \ref{fig:g11fit}, \ref{fig:allfits} and \ref{fig:silfits}) have either large $\chsq$s, indicating a poor correspondance between model and data, or $\chsq < 1$, which suggests that the model has overfit the data. In the latter case, it is clear that the SEDs of some objects can be fit with only two, or even one, dust component, rather than the four we use. However, the MCMC fitting approach accounts for degeneracies between model parameters. Figures \ref{fig:massratio} and \ref{fig:sizeratio} include all the information from the chain, so we are confident that the conclusions drawn from them are robust against overfitting issues.

For those SNRs with high $\chsq$ values, we note that these are typically driven by relatively small uncertainties on the far-IR fluxes. This is the wavelength range where background subtraction is most uncertain, as it contains the emission peak of cold ($\sim 20 \kel$) ISM dust, so there are likely to be significant additional systematic uncertainties not accounted for by our model. This wavelength range is also where the least well-constrained model parameters, such as the cold gas properties (Appendix \ref{sec:coldprop}), have the most impact on the model dust SEDs. {Modest variations in the assumed cold gas properties would almost certainly be able to obtain statistically-good fits to the data, without substantially altering our main conclusions.}

{We have assumed that the radiation field in both components is solely due to shock interactions, but Figure \ref{fig:sed} suggests that heating by the ISM radiation field could also be relevant, particularly for those SNRs located at small Galactocentric distances where the typical radiation field strength is higher \citep{mathis1983}. If we include an additional \citet{mathis1983} radiation field in the dust heating model, multiplied by a factor of five to approximate the stronger $5 \kpc$ field, the best-fit models (shown in Figure \ref{fig:g11ref}) have much lower small/large mass ratios in the cold component than those in Figure \ref{fig:g11fit} without the additional ISM field. However, the upper limit on this quantity (as given by the 84th percentile of the MCMC chain) is $0.36$ for carbon grains and $2.2$ for silicates, still consistent with values above the $0.16$ of the MRN size distribution (and the silicate median of $0.25$ is also higher than the ISM value). Depending on the details of the dust heating model, significant masses of small grains in the cold component may not be necessary to fit the observed SEDs, but they do not seem to be ruled out either, in contrast to the situation in the hot component.}

{Another implicit assumption in our models is that all IR flux comes from dust which has already been swept-up by the SNR. An alternative possibility is that the far-IR emission comes from dust grains ahead of the shock, which would explain the large dust masses relative to the inferred swept-up gas masses, and the grain size distributions being consistent with that in the ISM. We disfavour this explanation, as the shock radiation field should only penetrate a short distance into the ambient medium \citep{docenko2010}, and so the volume of dust preheated above ambient temperatures is quite small. The average density in this layer would then have to be orders of magnitude larger than our estimate above, to a somewhat implausible degree (see discussion in \citealt{priestley2021}). In any case, this possibility does not alter our conclusions about the grain sizes in the hot component, or that the mass of this component - in both gas and dust - is a negligible fraction of the total in the immediate surroundings of the SNR.}

\subsection{Implications}

The overall picture we find for our sample of SNRs is that the high-temperature, diffuse material contains a relatively small mass of large ($\gtrsim 0.1 \um$) grains, while dust in the denser, cooler shocked gas makes up virtually all the total swept-up dust mass, and {may} include a substantial {mass} of small ($\lesssim 10 \nm$) grains. This is {similar} to the situation in the Cassiopeia A reverse shock \citep{priestley2022}, and suggests that the physical processes affecting dust grains in shocks do not differ between the ISM and metal-enriched CCSN ejecta. {The observed distributions of grain sizes in the two gas components are suggestive of the two main processes responsible for processing dust grains in shocks:} small grains in the hot component are rapidly destroyed by sputtering; large grains in the cold component are efficiently converted into smaller grains via shattering in grain-grain collisions \citep{kirchschlager2019}.

{Although there is} {some} evidence for shattering ({large small-grain mass fractions}) in the dense, cool ejecta {(subject to the caveats discussed above)}, we find strict upper limits on the mass of small grains in the hot component for all SNRs. The hot component small/large mass ratios in Figure \ref{fig:sizeratio} suggest that either shattering is inefficient in this phase, or is at least subdominant to sputtering (i.e. newly-produced small grains are destroyed by sputtering on shorter timescales than they are produced by shattering of large grains). {Most theoretical studies of dust destruction \citep{jones1996,slavin2015,kirchschlager2022} assume ambient ISM densities of $\sim 0.1-1 \pcc$, similar to the hot component densities in Table \ref{tab:snrprop}, but find much more efficient shattering than appears to be the case in our SNRs (see e.g. the post-shock grain size distributions from \citealt{slavin2015}). These models may be overestimating the importance of grain-grain collisions in the processing of shocked ISM dust.}

{The severe mismatch between the estimated total swept-up dust masses, if the ambient ISM densities are those derived from X-ray measurements, and the observed present-day dust masses in Table \ref{tab:mass}, effectively requires that the total (gas plus dust) mass of the hot component is a small fraction of that swept up by the SNRs. Making the conservative assumption that no dust has been destroyed, the implied swept-up {(or soon to be swept-up)} gas masses for a DTG ratio of $0.01$ are (with the exception of G29) hundreds of times larger than those of the SNR hot components (estimated using the gas densities and volumes in Table \ref{tab:snrprop}). Any post-shock reduction in dust mass, or a hot component filling factor lower than unity, will make this discrepancy larger. When considering the overall effect of the SNR on the surrounding ISM, what occurs in (or to) the hot component - the only phase typically modelled by theoretical work - is effectively negligible.

As the gas temperatures in {the cold component} are unlikely to be high enough for thermal sputtering\footnote{{Kinetic sputtering may be effective regardless of gas temperature for a sufficiently strong shock, although in higher-density gas the shock strength is also necessarily reduced, while the coupling between gas and grain motions is increased. We would thus expect the kinetic sputtering rate to also be lower in the cold component.}} to be effective {($> 10^5 \kel$; \citealt{biscaro2016})}, and grain collisions themselves mostly reprocess rather than destroy dust \citep{kirchschlager2019}, it is conceivable that {a significant fraction} of the dust in this component has survived being shocked. {It is generally thought that the majority of the mass in the ISM is contained in much denser, colder phases than those typically investigated by models of dust destruction \citep{mckee1977,jones2011}. There is observational evidence, in some cases, for a large fraction of the total SN momentum and energy going into these phases \citep[e.g.][]{cosentino2022}, rather than the more diffuse ISM. If the destruction efficiency in the dense ISM is {in fact} lower, it seems likely that models assuming a uniform, low-density ISM are overestimating the rate of dust destruction, {particularly if grain shattering in the low-density ISM is also being overestimated (which may have a huge impact on the destruction efficiency; \citealt{kirchschlager2022})}.}

\section{Conclusions}

We have modelled the IR dust emission from the shocked ISM for a sample of seven SNRs, taking into account the multiphase nature of the shocked material, and the uncertain (and likely non-standard) grain properties. We find consistent results {across the} sample: grains located in the {cooler} ($\sim 1000 \kel$) shocked gas make up $> 90\%$ of the total surviving swept-up dust mass; only large ($\gtrsim 0.1 \um$) grains have survived in the hot ($> 10^6 \kel$) phase of the shocked ISM; {grain size distributions in the colder phase are consistent with those in the ISM, or possibly even biased towards small ($\lesssim 10 \nm$) grains}. We {suggest} that this indicates efficient sputtering in the hot phase, and {either} efficient shattering {or generally inefficient dust processing} in the cold phase.

The lack of evidence for grain shattering in the hot, {diffuse swept-up ISM} is contrary to models of dust destruction in shocks. {Most theoretical {predictions of the} dust destruction efficiency {in SNRs} assume a uniform, low-density ambient medium. The multi-phase nature of the observed SNRs, with a very small fraction of the total dust mass in the low-density material, suggests that these predicted values may be significantly overestimating the dust destruction efficiency of SNe.}

\section*{Acknowledgements}

{We are grateful to Florian Kirchschlager for comments on a draft version of this paper.} FDP is supported by a consolidated grant (ST/K00926/1) from the Science and Technology Facilities Council. HC and HLG acknowledge support from the European Research Council (ERC) grant COSMICDUST ERC-2014-CoG-647939. IDL acknowledges support from ERC starting grant 851622 DustOrigin. MJB acknowledges support from the ERC grant SNDUST ERC-2015-AdG-694520.

\section*{Data Availability}

The data underlying this article will be made available upon request.




\bibliographystyle{mnras}
\bibliography{hotdust}


\appendix

\section{SNR dust fluxes}
\label{sec:snrsed}

Table \ref{tab:flux} lists the IR fluxes used in the dust modelling for each SNR in our sample.

\begin{table*}
  \centering
  \caption{IR SEDs for our SNR sample, with fluxes given in Jy and filter effective wavelength in $\um$. Data for G11, G27 and G29 are taken from \citet{priestley2021}; G43, G340 and G349 from \citet{chawner2020}; and the Tornado from \citet{chawner2020b}.}
  \begin{tabular}{ccccccccc}
    \hline
    SNR & IRAC $8$ & WISE $12$ & MIPS $24$ & PACS $70$ & PACS $160$ & SPIRE $250$ & SPIRE $350$ & SPIRE $500$ \\
    \hline
    G11 & - & $<6.2$ & $26.0 \pm 3.1$ & $124.8 \pm 22.4$ & $176.8 \pm 85.6$ & $65.0 \pm 52.1$ & $33.4 \pm 21.9$ & $15.3 \pm 7.7$ \\
    G27 & - & $<1.2$ & $13.0 \pm 0.2$ & $25.6 \pm 21.6$ & $124.7 \pm 70.0$ & $94.4 \pm 29.5$ & $47.1 \pm 13.2$ & $16.6 \pm 4.5$ \\
    G29 & - & $<1.9$ & $10.0 \pm 2.1$ & $101.5 \pm 39.1$ & $16.1 \pm 112.4$ & $11.4 \pm 47.4$ & $1.3 \pm 17.3$ & $2.2 \pm 8.2$ \\
    Tornado & $<1.66$ & - & $4.3 \pm 0.2$ & $164.5 \pm 11.5$ & $151.2 \pm 10.6$ & $63.2 \pm 3.5$ & $25.9 \pm 1.4$ & $8.2 \pm 0.5$ \\
    G43 & $<61.0$ & - & $80.1 \pm 6.0$ & $744.5 \pm 61.2$ & $283.6 \pm 135.2$ & $74.4 \pm 113.7$ & $20.9 \pm 53.8$ & $5.3 \pm 23.1$ \\
    G340 & $<24.5$ & - & $23.2 \pm 7.8$ & $220.6 \pm 29.4$ & $755.0 \pm 117.8$ & $506.5 \pm 85.1$ & $207.5 \pm 36.7$ & $80.2 \pm 14.1$ \\
    G349 & $<12.6$ & - & $47.8 \pm 3.5$ & $588.4 \pm 41.5$ & $456.5 \pm 32.8$ & $158.8 \pm 9.1$ & $60.1 \pm 3.5$ & $19.6 \pm 1.2$ \\
    \hline
  \end{tabular}
  \label{tab:flux}
\end{table*}

\section{Cold gas properties}
\label{sec:coldprop}

{For collisional heating in the cold component, we have assumed gas properties derived from H$_2$ line observations by \citet{priestley2021} for G11. There is no guarantee that these properties are appropriate for the other SNRs in our sample. Figure \ref{fig:coltest} shows the influence of the assumed cold component properties on the resulting grain fluxes. We consider three cases: the G11 properties ($\nh = 1000 \pcc$, $\nel = 0.1 \pcc$, $T = 5000 \kel$); radiative heating only, with collisional heating turned off; and $\nh = \nel = 10 \pcc$ with $T = 10^4 \kel$, representing more diffuse, fully-ionised gas. The emissivity of $0.1 \um$ grains is almost completely unaffected by the assumed cold component gas properties, as the grain heating is dominated by the radiation field in all cases. For $5 \nm$ grains, there is a significant difference between the case with no collisional heating and the two cases where it is included, but the grain SEDs are very similar for the two sets of gas properties we consider.}

{While investigating the full three-dimensional collisional heating parameter space ($\nh, \nel, T$) is beyond the scope of this paper, it appears that for a range of post-shock dense gas properties, dust SEDs are similar enough to be identical for our purposes. Even if we neglect collisional heating entirely, our main conclusions are qualitatively unchanged. Figure \ref{fig:g11radfit} shows best-fit dust masses for G11, with the cold component only heated by the radiation field. While the dust masses differ from those in Figure \ref{fig:g11fit}, regardless of grain composition we still find that the majority of the dust mass is in the cold component, that the mass of small grains is comparable to that of large grains in this component, and that grains in the hot component (if present) must be large.}

\begin{figure}
  \centering
  \subfigure{\includegraphics[width=\columnwidth]{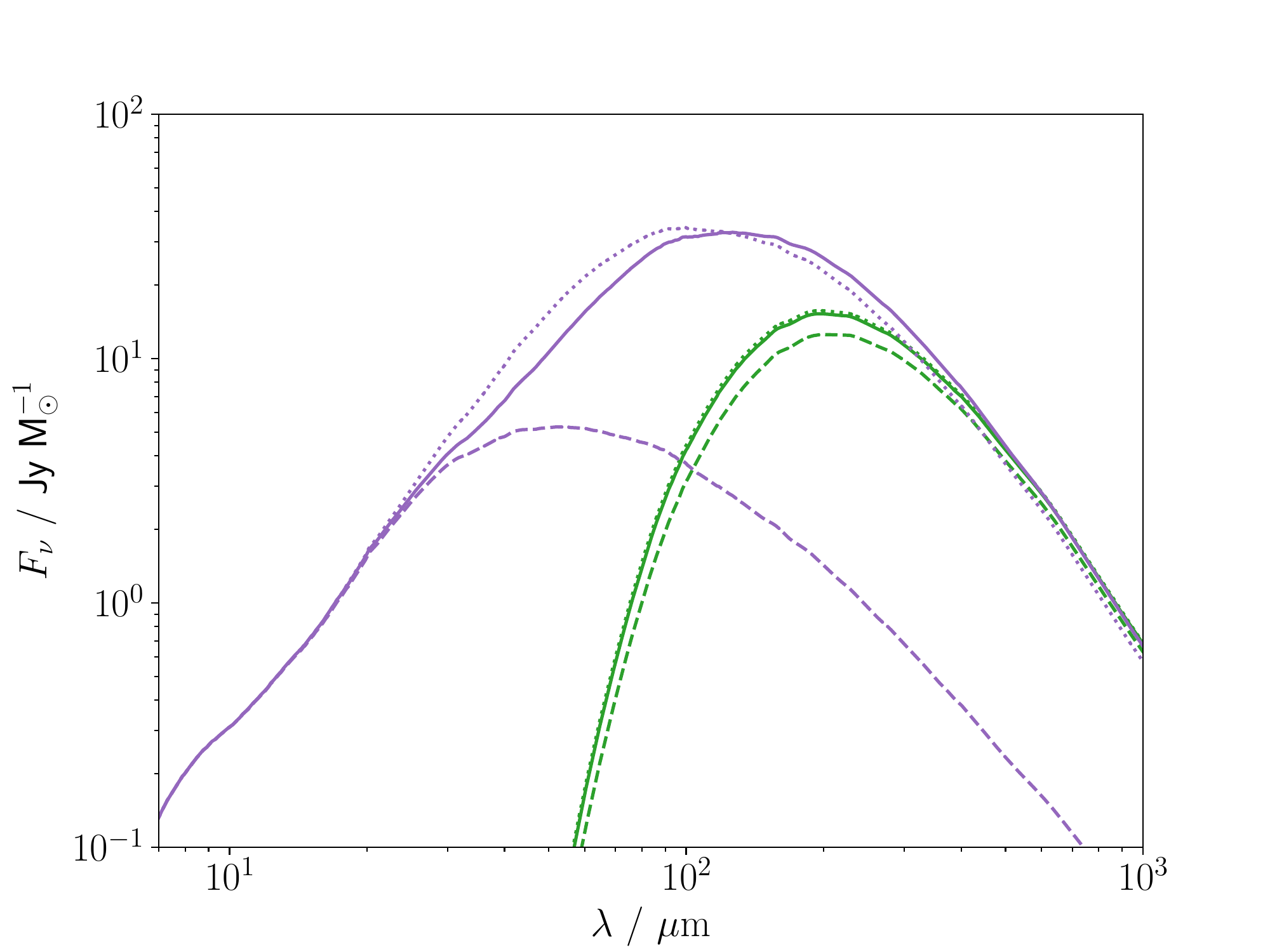}}\quad
  \caption{Flux per unit mass for cold/large (green) and cold/small (purple) carbon grains in G11. Solid lines have gas properties derived from H$_2$ observations ($\nh = 1000 \pcc$, $\nel = 0.1 \pcc$, $T = 5000 \kel$), dashed lines are heated solely by the radiation field, and dotted lines assume $\nh = \nel = 10 \pcc$, $T = 10^4 \kel$.}
  \label{fig:coltest}
\end{figure}

\begin{figure*}
  \centering
  \subfigure{\includegraphics[width=\columnwidth]{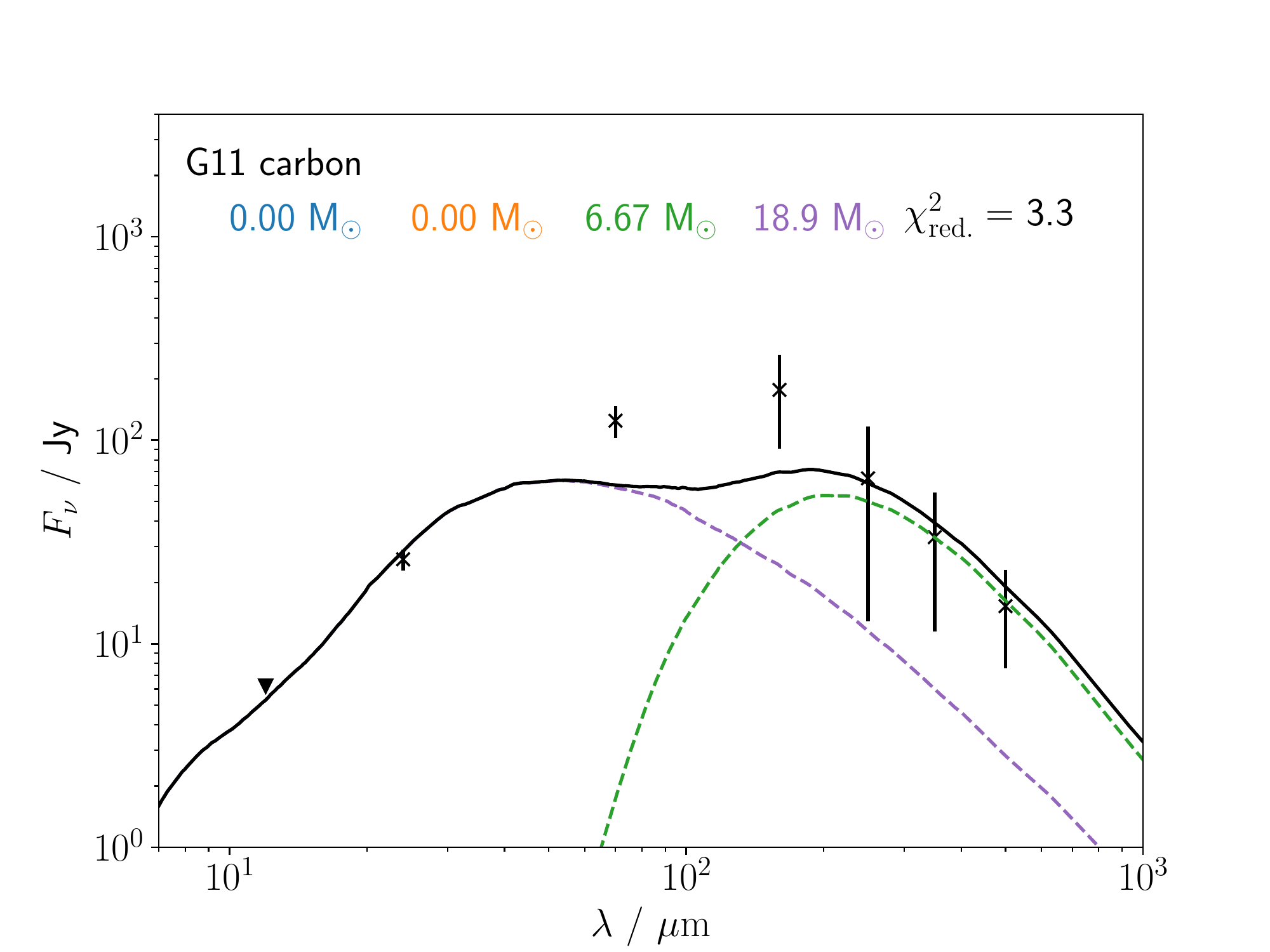}}\quad
  \subfigure{\includegraphics[width=\columnwidth]{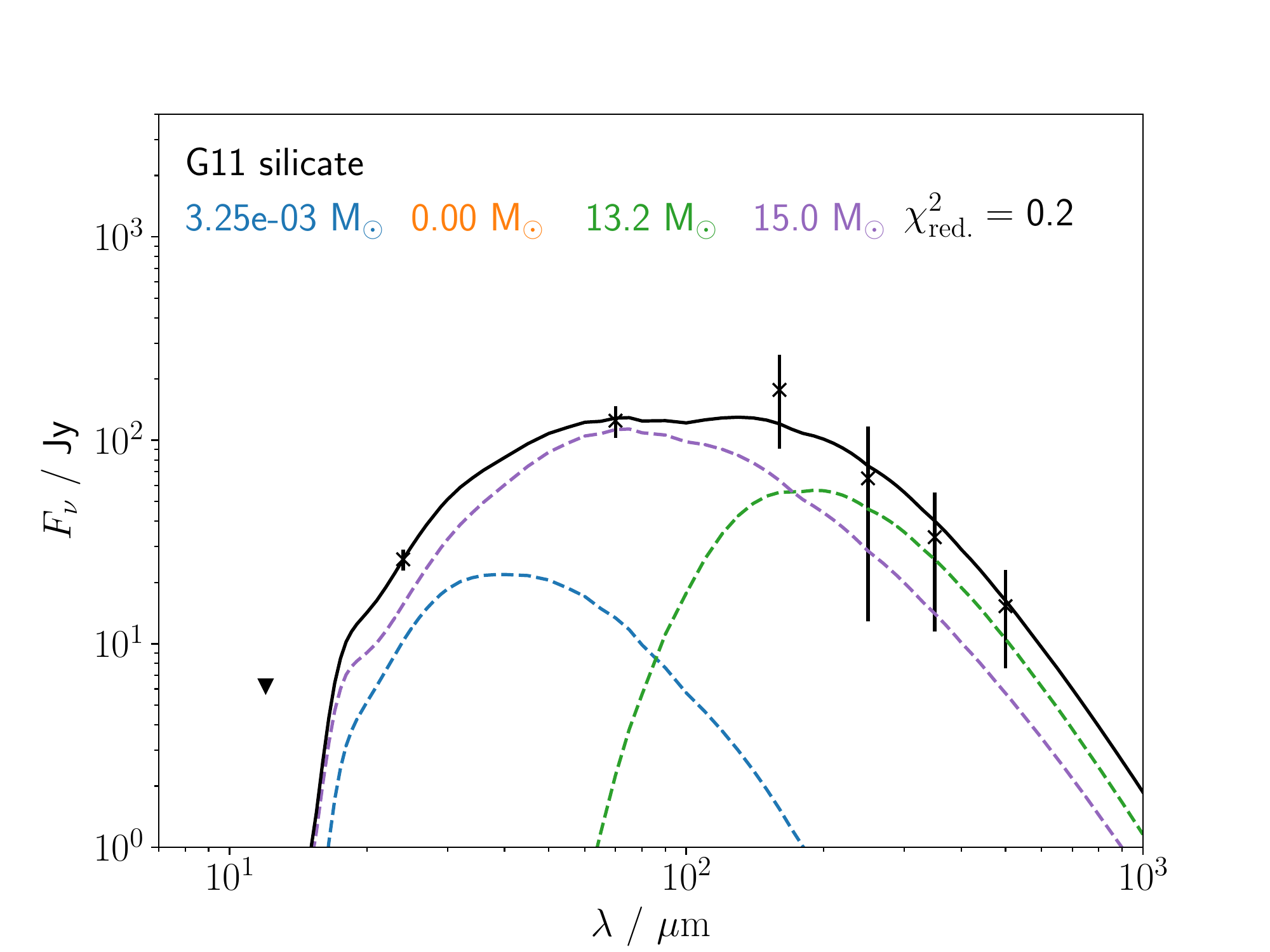}}
  \caption{Best-fit dust SEDs for G11 {with no collisional heating by the cold component} - data (black crosses), total model SED (black line), and individual component SEDs: hot/large (blue); hot/small (orange); cold/large (green); cold/small (purple). {\it Left:} carbon grains. {\it Right:} silicates.}
  \label{fig:g11radfit}
\end{figure*}

\section{Grain size distributions}
\label{sec:sizedist}

\begin{figure*}
  \centering
  \subfigure{\includegraphics[width=\columnwidth]{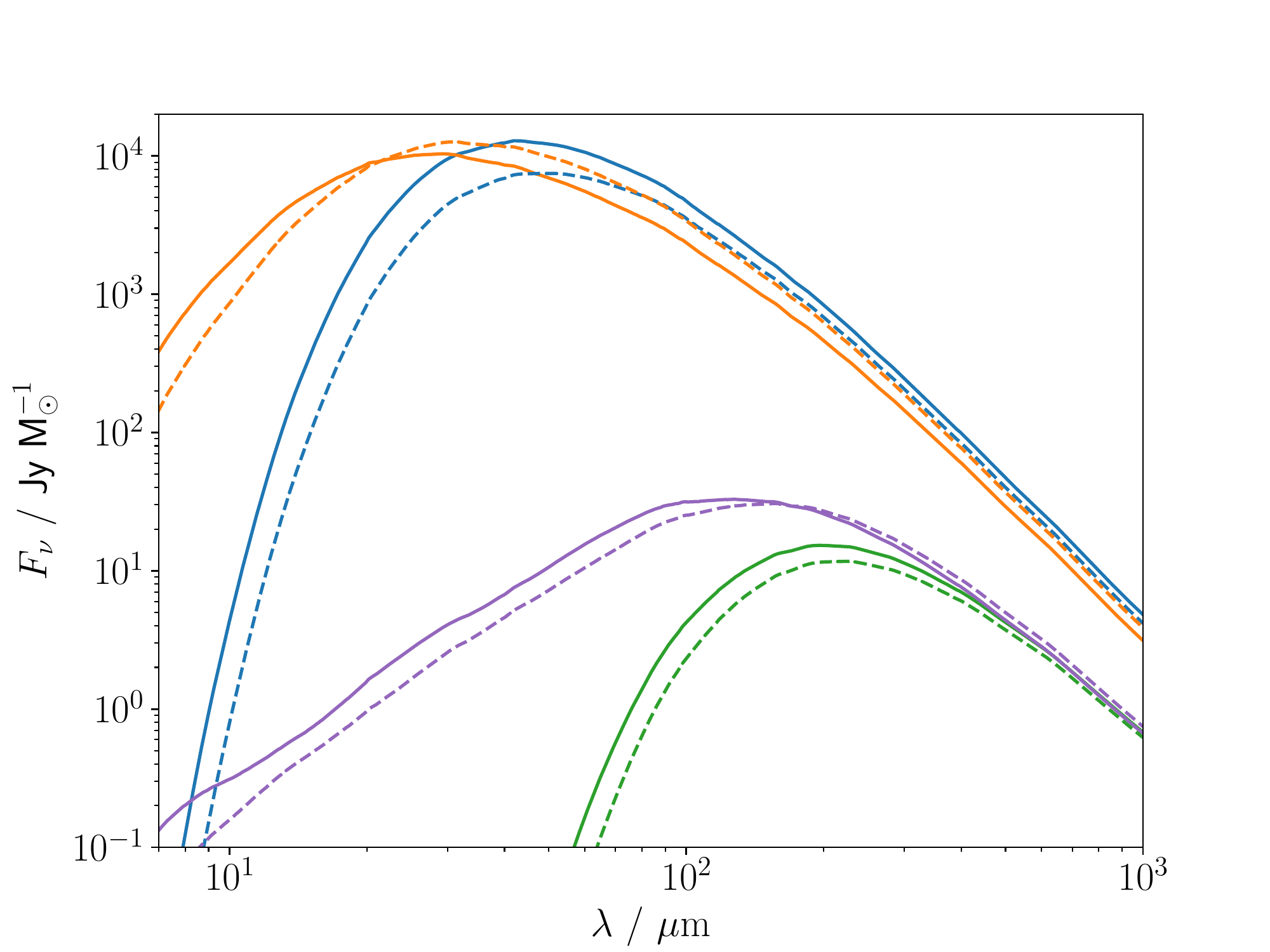}}\quad
  \subfigure{\includegraphics[width=\columnwidth]{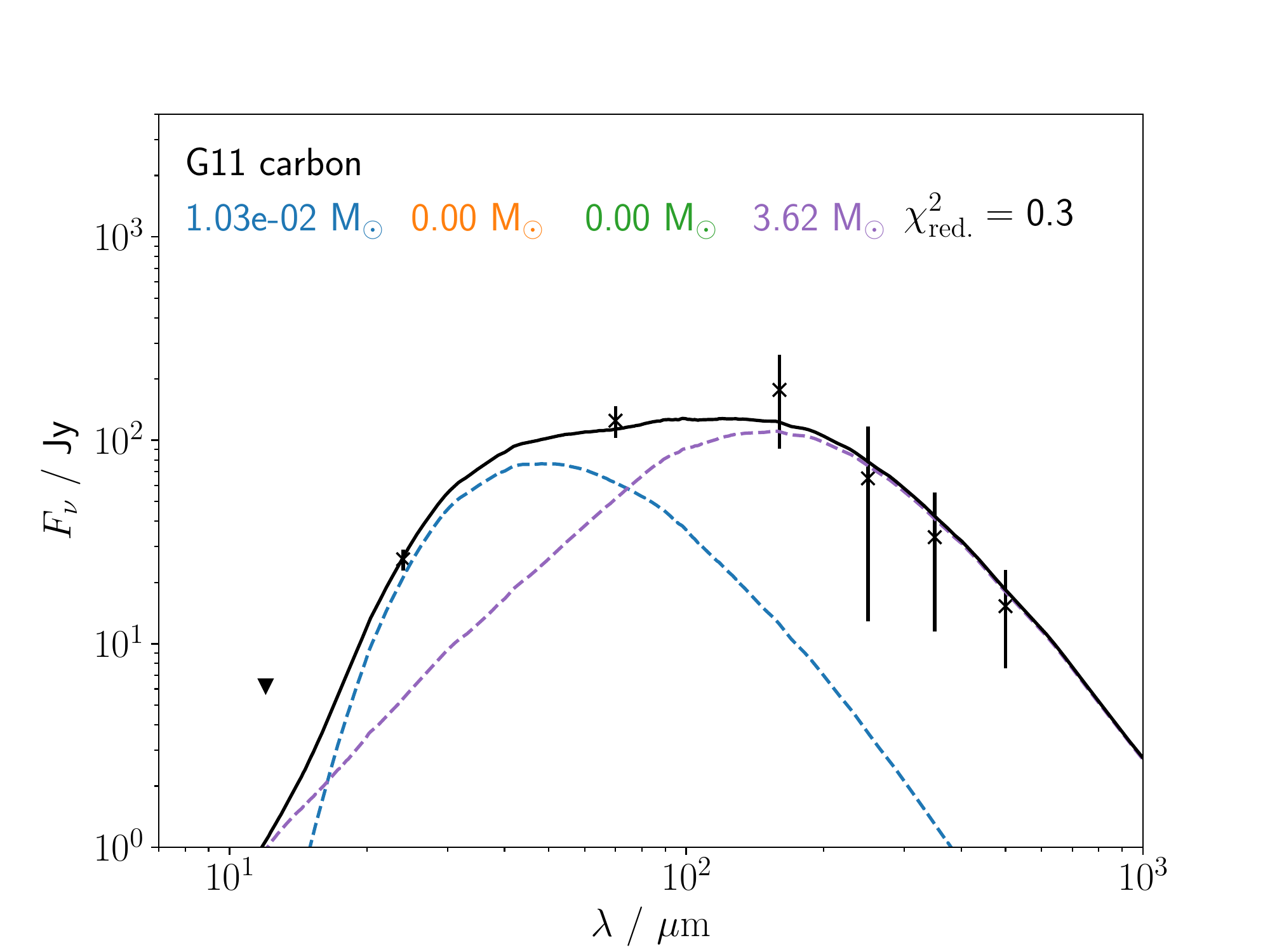}}
  \caption{{\it Left:} flux per unit mass for carbon grains in G11: hot/large (blue); hot/small (orange); cold/large (green); cold/small (purple). Solid lines are for single grain sizes, dashed lines for size distributions. {\it Right:} best-fit dust SED for G11 with a distribution of grain sizes - data (black crosses), total model SED (black line), and individual component SEDs: hot/large (blue); hot/small (orange); cold/large (green); cold/small (purple).}
  \label{fig:g11dist}
\end{figure*}

{Our choice of $0.1 \um$ to represent large grains, and $5 \nm$ to represent small ones, is somewhat arbitrary, although informed by the sizes typically thought to be present in the ISM. Figure \ref{fig:g11dist} shows the impact of using a size {\it distribution} for each component, rather than a single representative grain size, using carbon grains in G11 as an example. We replace small and large grains with power-law distributions, with an MRN exponent of $-3.5$, spanning the ranges $5-10 \nm$ and $0.1-0.3 \um$ respectively. Both the SEDs of the individual dust components, and the results of the MCMC fit to the G11 data, are affected by at most a factor of a few. This is a minor source of uncertainty compared to others in our modelling procedure, and is not sufficient to qualitatively change our conclusions to any significant extent.}

\section{Silicate grain SED fits}
\label{sec:silfit}

\begin{figure*}
  \centering
  \subfigure{\includegraphics[width=\columnwidth]{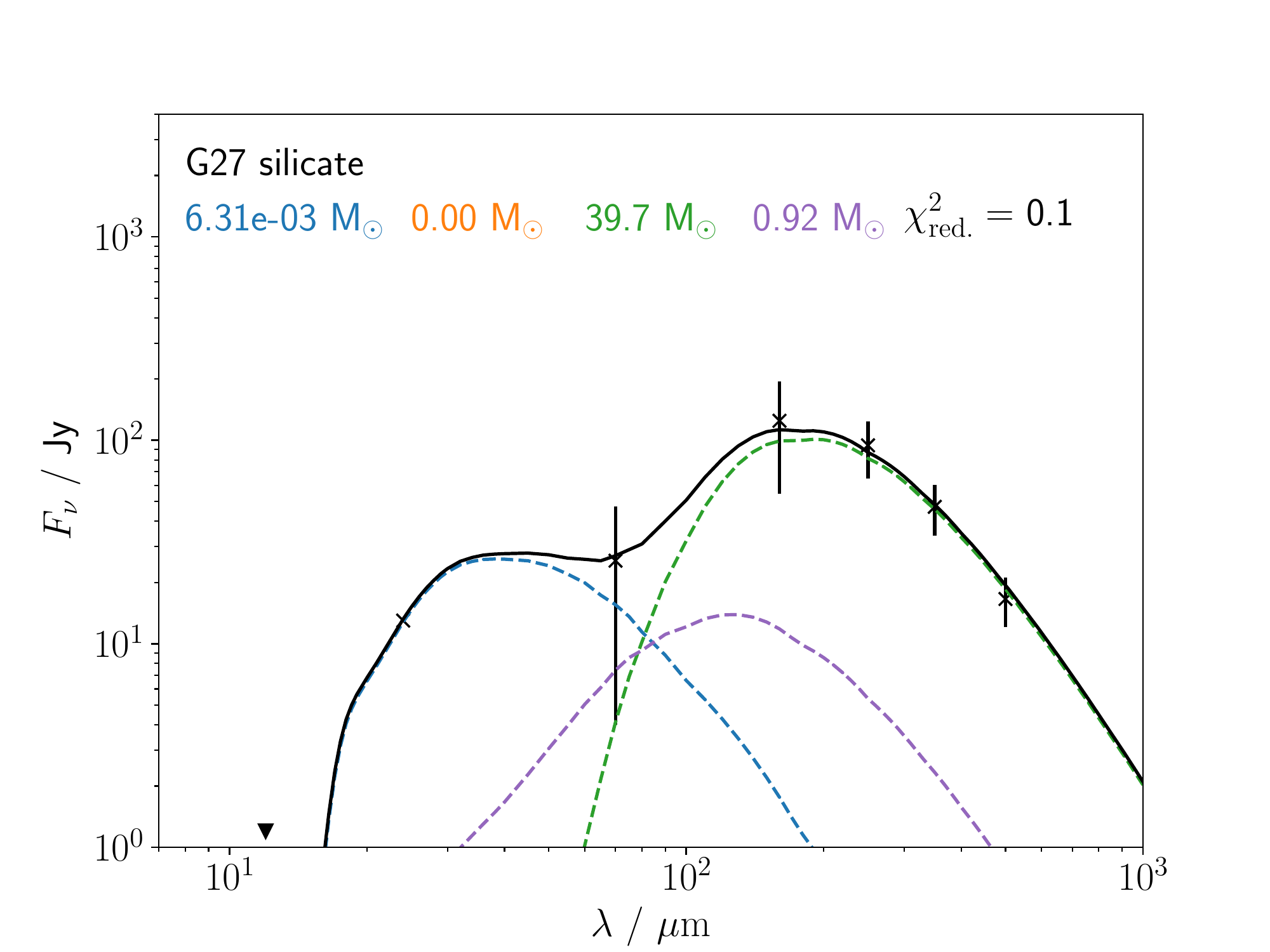}}\quad
  \subfigure{\includegraphics[width=\columnwidth]{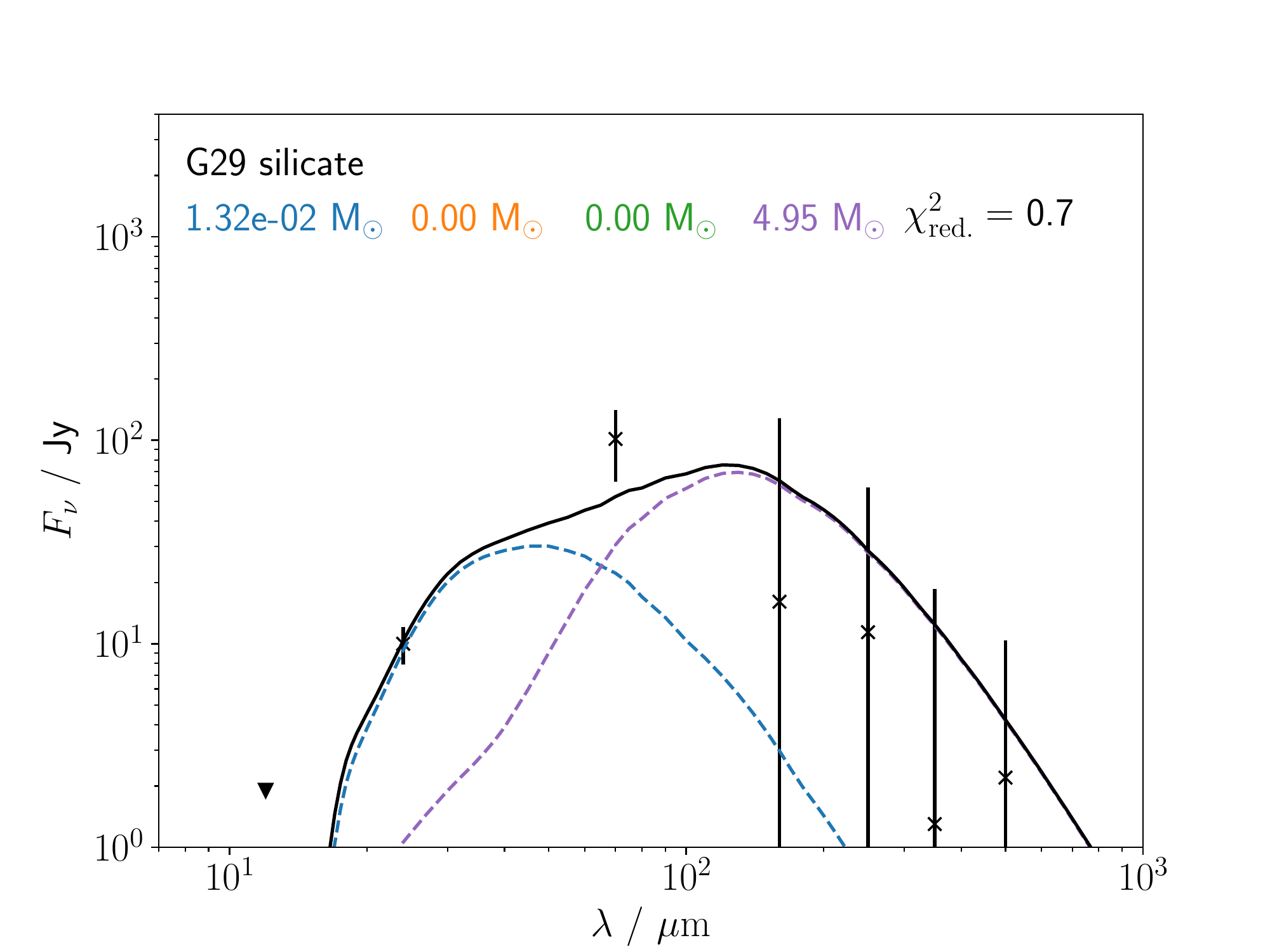}}\\
  \subfigure{\includegraphics[width=\columnwidth]{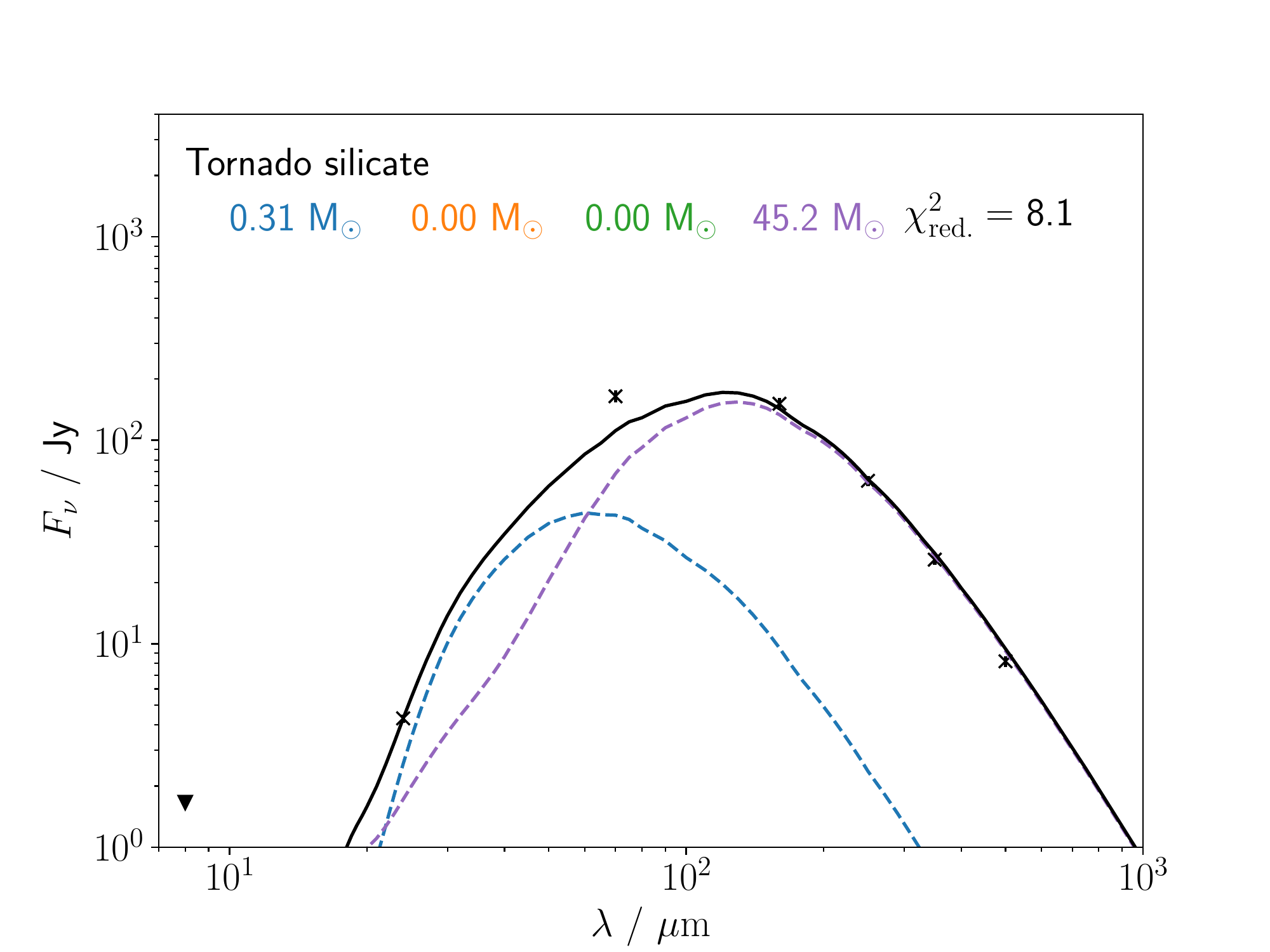}}\quad
  \subfigure{\includegraphics[width=\columnwidth]{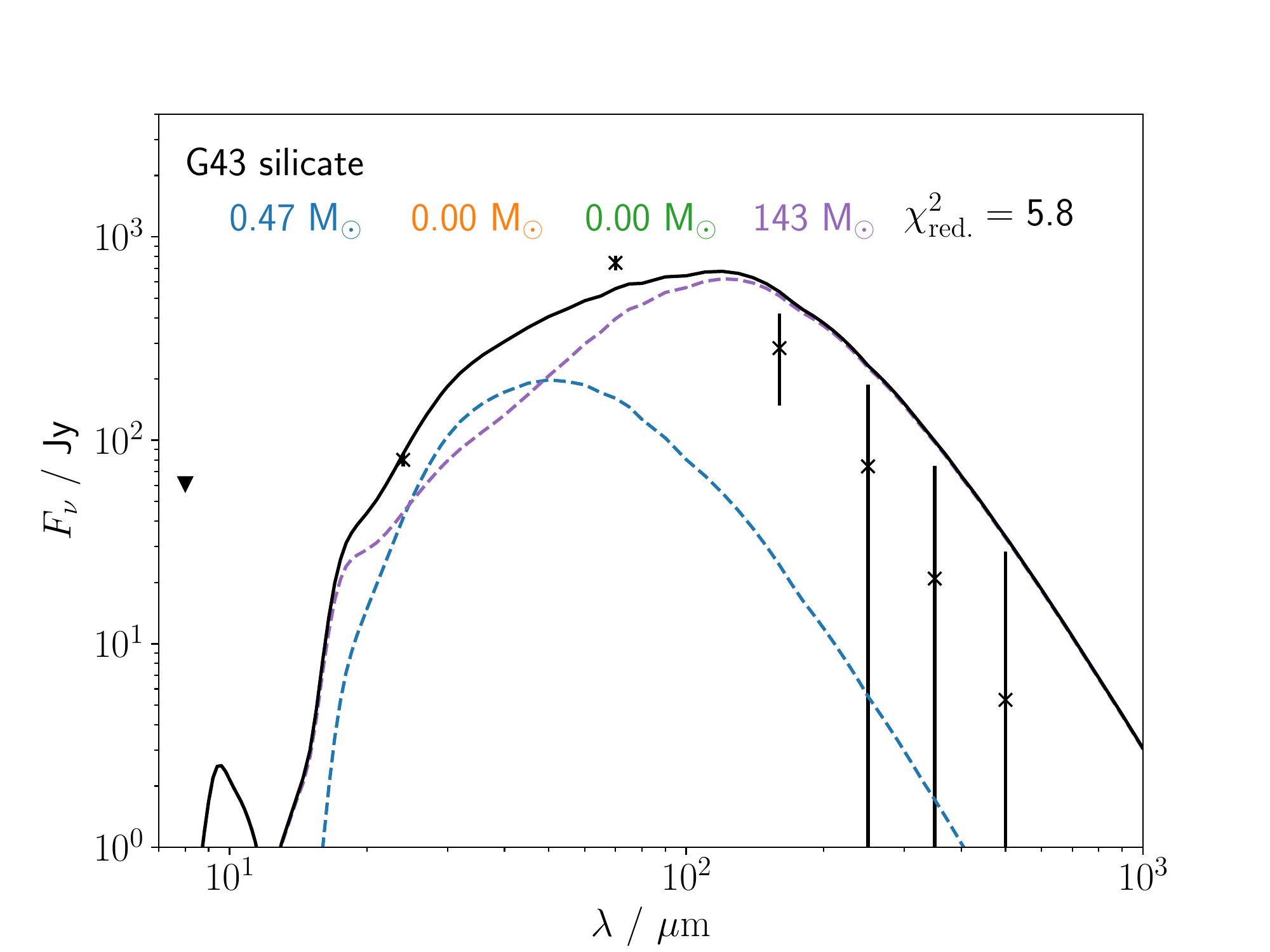}}\\
  \subfigure{\includegraphics[width=\columnwidth]{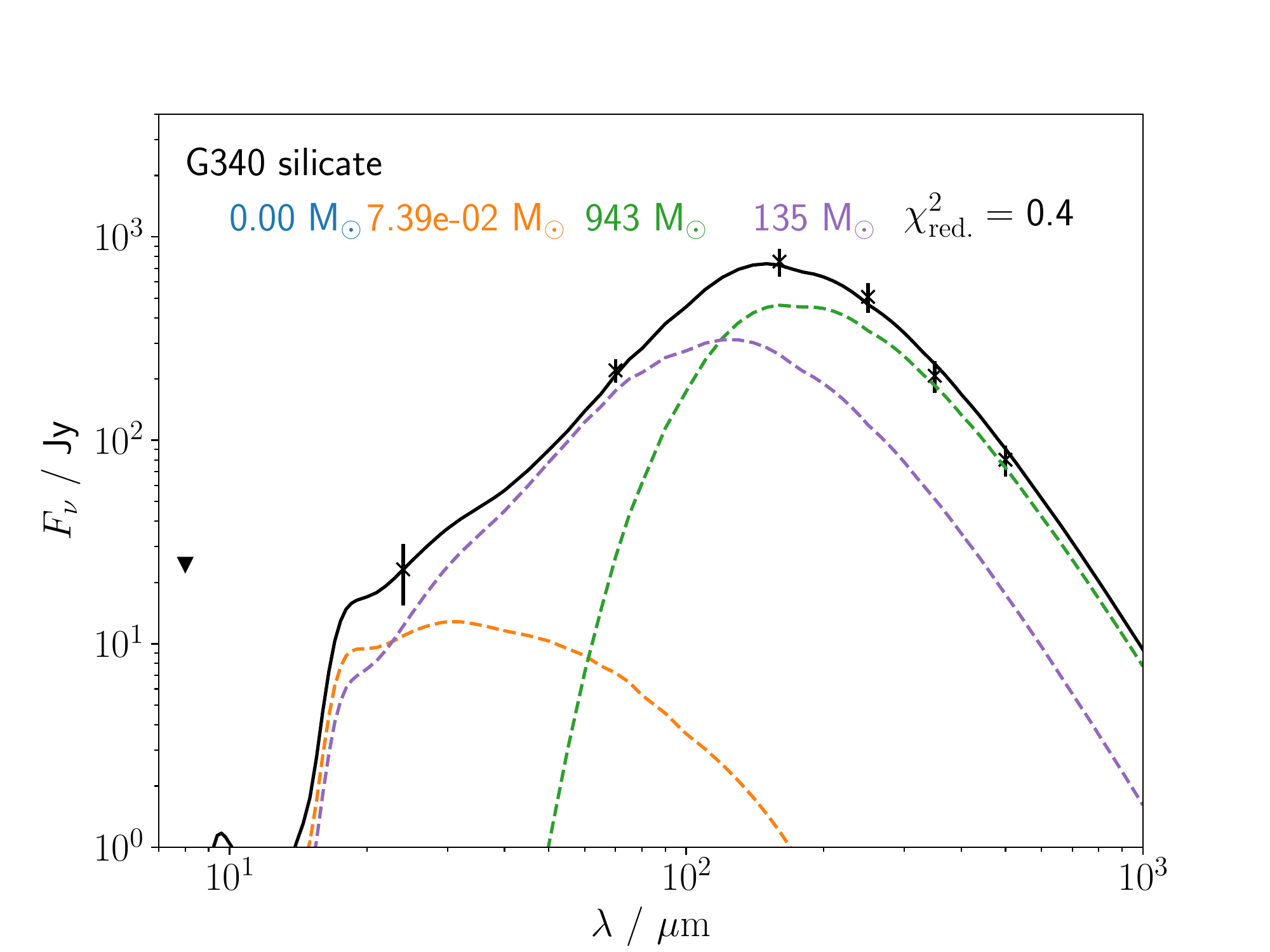}}\quad
  \subfigure{\includegraphics[width=\columnwidth]{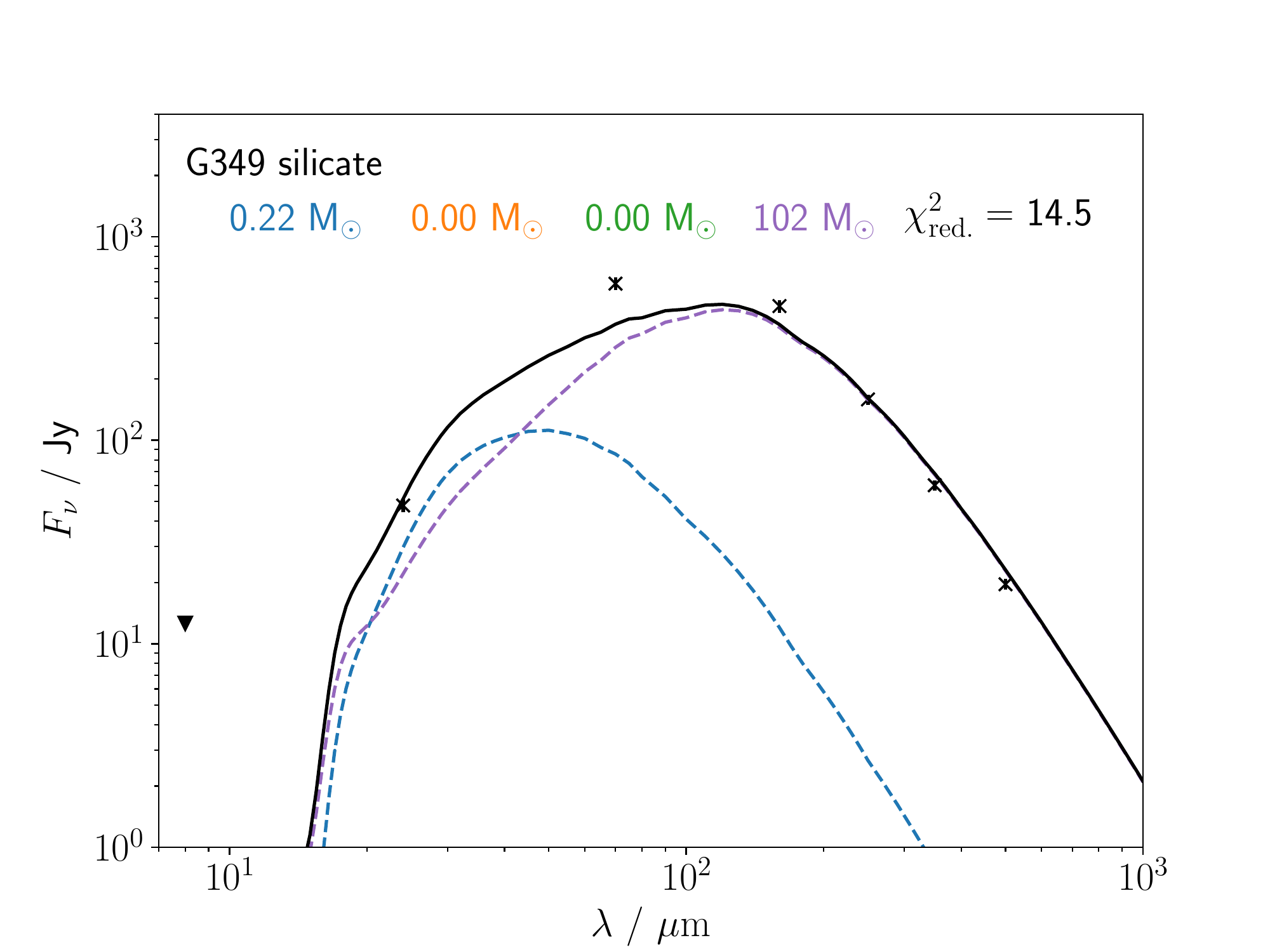}}\\
  \caption{Best-fit dust SEDs for silicate grains - data (black crosses), total model SED (black line), and individual component SEDs: hot/large (blue); hot/small (orange); cold/large (green); cold/small (purple). {\it Top left:} G27. {\it Top right:} G29. {\it Middle left:} Tornado. {\it Middle right:} G43. {\it Bottom left:} G340. {\it Bottom right:} G349.}
  \label{fig:silfits}
\end{figure*}

Figure \ref{fig:silfits} shows the best-fit silicate grain SEDs for our sample of SNRs, excluding G11, which is presented in Figure \ref{fig:g11fit}. {While the different grain composition results in non-trivial changes to both the total dust mass and its distribution between components, we find identical qualitative results across the sample of SNRs as for carbon grains: the hot component is typically a small fraction of the total dust mass, and rarely (if ever) contains any substantial mass of small grains, whereas the cold component requires a non-negligible mass of small grains.}


\bsp	
\label{lastpage}
\end{document}